%% file: paper_v1.01mf.tex
\documentclass[useAMS,usenatbib]{mnras}
\usepackage{verbatim}
\usepackage{placeins}
\usepackage{float}
\usepackage[utf8]{inputenc}
\usepackage{mathtools}
\usepackage{amsmath}
\usepackage{graphicx}
\usepackage{multirow}
\usepackage{multicol}
\usepackage{ulem}
\newcommand{\Mlum}{\ensuremath{M_\mathrm{L}}}

\newcommand{\Mdyn}{\ensuremath{M_\mathrm{L+D}}}
\newcommand{\Mtot}{\ensuremath{M_\mathrm{dyn}}}
\newcommand{\Mjam}{\ensuremath{M_\mathrm{JAM}}}
\newcommand{\Mlim}{\ensuremath{M_\mathrm{lim}}}
\newcommand{\Medyn}{\ensuremath{M_{\rm e}^{\rm dyn}}}
\newcommand{\Mhalfdyn}{\ensuremath{M_{1/2}^{\rm dyn}}}

\newcommand{\Mstars}{\ensuremath{M_\mathrm{stars}}}
\newcommand{\Mdark}{\ensuremath{M_\mathrm{dark}}}

\newcommand{\Reff}{\ensuremath{R_\mathrm{e}}}
\newcommand{\rhalflight}{\ensuremath{r_\mathrm{1/2}}}
\newcommand{\RShen}{\ensuremath{R_\mathrm{Shen}(\Mlum)}}
\newcommand{\CShen}{\ensuremath{C_\mathrm{Shen}}}
\newcommand{\kpc}{\ensuremath{\mathrm{kpc}}}
\newcommand{\rgrav}{\ensuremath{r_\mathrm{g}}}
\newcommand{\rcutoff}{\ensuremath{r_\mathrm{cutoff}}}
\newcommand{\aH}{\ensuremath{a_\mathrm{H}}}

\newcommand{\sigmae}{\ensuremath{\sigma_\mathrm{e}}}

\newcommand{\fdm}{\ensuremath{f_\mathrm{DM}}}

\newcommand{\nsers}{\ensuremath{n}}

\newcommand{\kpar}{\ensuremath{K}}
\newcommand{\klumnoDM}{\ensuremath{K_\star^{\rm noDM}}}
\newcommand{\khalfdyn}{\ensuremath{K_{1/2,\mathrm{dyn}}}}
\newcommand{\klum}{\ensuremath{K_\star}}
\newcommand{\klim}{\ensuremath{K_\mathrm{lim}}}

\newcommand{\kkin}{\ensuremath{K_{\rm kin}}}
\newcommand{\kstr}{\ensuremath{K_{\rm str}}}

\newcommand{\EKIN}{\ensuremath{E_\mathrm{K}}}
\newcommand{\EPOT}{\ensuremath{E_\mathrm{P}}}

\title[Non-homology of ellipticals grown by dry mergers]
{Dynamical masses and non-homology of massive elliptical galaxies grown by dry mergers}
\author[M. Frigo and M. Balcells]{M. Frigo$^{1}$\thanks{E-mail:
mfrigo@mpa-garching.mpg.de} and M. Balcells$^{2,3,4}$\\
$^{1}$Max Planck Institute for Astrophysics, Garching, Germany\\
$^{2}$Isaac Newton Group of Telescopes, E-38700 Santa Cruz de La Palma, La Palma, Spain\\
$^{3}$Instituto de Astrof\'\i sica de Canarias (IAC), E-38205 La Laguna, Tenerife, Spain\\
$^{4}$Universidad de La Laguna, Departamento de Astrof\'{\i}sica, E-38206 La Laguna, Tenerife, Spain}

\begin{document}

\date{Accepted 2017 April 05. Received 2017 April 05; in original form 2016 October 30}

\pagerange{\pageref{firstpage}--\pageref{lastpage}} \pubyear{2017}

\maketitle

\label{firstpage}

\begin{abstract}
We study whether dry merger-driven size growth of massive elliptical galaxies depends on their initial structural concentration, and analyse the validity of the homology hypothesis for virial mass determination in massive ellipticals grown by dry mergers.
High-resolution simulations of a few realistic merger trees, starting with compact progenitors of different structural concentrations (S\'ersic indices $n$), show that galaxy growth has little dependence on the initial S\'ersic index (larger $n$ leads to slightly larger size growth), and depends more on  other particulars of the merger history. We show that the deposition of accreted matter in the outer parts leads to a systematic and predictable breaking of the homology between remnants and progenitors, which we characterize through the evolution, during the course of the merger history, of  virial coefficients $\kpar \equiv G M / \Reff\, \sigmae^2$ associated to the most commonly-used dynamical and stellar mass parameters. The virial coefficient for the luminous mass, \klum\ , is $\sim$50 per cent larger at the $z \approx 2$ start of the merger evolution than in $z=0$ remnants.   
Ignoring virial evolution leads to biased virial mass estimates. 
We provide \kpar\ corresponding to a variety of dynamical and stellar mass parameters, and provide recipes for the dynamical determination of galaxy masses. 
For massive, non-compact ellipticals, the popular expression $M = 5\, \Reff \,\sigmae^2\,/\,G$ underestimates the  dynamical mass within the luminous body by factors of up to 4; it instead provides an approximation to the total \textit{stellar} mass with smaller uncertainty than current stellar population models. 
\end{abstract}


\section{Introduction}
\label{Sec:Introduction} 

In the last decade various observations have shown that for a given stellar mass galaxies at redshift $z \sim 1.5-2.5$ were much more compact than today \citep{daddi2005,trujillo2006,trujillo2007,vandokkum2008,buitrago2008, vdw2014}.  Galaxies with a stellar mass of $M \sim 10^{12}\,{\rm M}_{\odot}$ had effective radii of $\Reff \sim 1 \, \kpc$, while today's galaxies of that mass have an effective radius of the order of $5 \, {\rm kpc}$. This means that the overall density of high redshift ellipticals was more than 50 times higher. There have been several attempts to dismiss these values as an observational error, such as a strong morphological K-correction or the presence of an active galactic nucleus \citep{daddi2005}, but later observations \citep{trujillo2006} disproved these alternative explanations, and the existence of these compact galaxies at high redshift is now well enstablished. Their remarkable density, coupled with their low star formation rates and gas content earned them the nickname `red nuggets' \citep{damjanov2009}. More recent high redshift observations showed that a population of similarly compact but star-forming galaxies existed at redshift $z>2$, suggesting that red nuggets formed by gravitational collapse of gas into a small region followed by intense star formation and rapid quenching \citep{dekel2014, vd2015}. In the local universe however, galaxies of this compactness have been shown to be extremely rare \citep{bernardi2006,trujillo2009,ferremateu2012,pda2016}. This suggests that between $z=2$ and $z=0$ they have probably evolved into the more extended elliptical galaxies we see in the local universe. \newline
The mechanism which caused this growth in size is still not completely clear, but the most widely supported explanation holds that the size growth of these galaxies is produced mainly by major and/or minor dry mergers. Dry mergers are mergers in which the gas content of the interacting galaxies is negligible, which is a reasonable approximation after the massive, compact galaxy is formed. This hypothesis has received support from both observations and numerical simulations. On the observational side it has been shown that elliptical galaxies have likely grown inside-out, accreting matter around their original dense cores \citep{vandokkum2010,trujillo2011,cooper2012}. In many cases the dense cores of local massive ellipticals have the same mass and density of the compact galaxies at high redshift \citep{vd2014}, indicating that this growth process does not alter the matter distribution of the original galaxy, but mostly extends it with new material. \citet{delarosa2016} showed that the number density of galaxies hosting a compact core at z=0 is compatible with the number density of compact ellipticals at high redshift, at least when also considering disc-like galaxies with a compact core. 
And while relic compact galaxies in the local universe are extremely rare, a merger-driven growth would also explain their relatively higher frequency in cluster environments \citep{poggianti2013,pda2016}, where mergers are less common due to the larger relative speeds of galaxies. 
On the computational side, many studies have shown that dry mergers can produce the observed size growth \citep{naab2009} and in particular that minor mergers are the most efficient at increasing the size of the galaxy \citep{hilz2012}. This two-phase formation process of massive ellipticals has also been succesfully recreated in cosmological simulations \citep{oser2010, oser2012}, with the expected size growth rate. Even though some computational studies argued that additional physical processes must be added to the pure dry merger scenario in order to explain the observed growth \citep{nipoti2009}, especially at high redshift \citep{cimatti2012}, it is by now accepted that dry mergers are the main mechanism driving this evolution process. 
\newline
If massive elliptical galaxies grew via minor mergers, in which accreted matter settles mostly in the outer parts without affecting the core, then galaxies will acquire a two-component, core+envelope structure. Homology, the property that two galaxies transform into each other by direct scaling in size and/or mass, would be broken along the merger growth.  
Validating homology is relevant for sorting out the unphysical result that for many high-redshift massive galaxies, and for some nearby compact ellipticals,  dynamical masses \Mtot\ from the virial theorem are lower than luminous masses \Mlum\ from stellar-population modelling \citep{stockton2010, martinezmanso2011, ferremateu2012, pda2014}. 
These authors obtain \Mtot\ from the formula:
\begin{equation}
\label{eqn:kpardef}
\Mtot = \kpar\,\Reff\,\sigmae^2\,/\,G, 
\end{equation}
\noindent (where \Reff\ is the effective radius and \sigmae\ is the luminosity-weighted velocity dispersion within \Reff; see Section~\ref{sec:virialeqns}), 
citing \citet{cappellari2006} to choose $\kpar\approx 5$ and assuming homology in order to apply this value of \kpar\ to all their galaxies. 
To explain the unphysical $\Mlum > \Mtot$ result,
\citet{martinezmanso2011}  explored whether stellar-population masses were being over-estimated 
,  whereas
\citet{pda2014,pda2015} 
 studied observational clues to the assumption of homology. The latter authors argued that the mass--size--velocity dispersion data of massive galaxies over different cosmological epochs, as well as at a given $z$, are incompatible with homology. \citet{pda2014} show that the ratio $\Mlum/\Mtot$ scales with the compactness index of the spatial mass distribution \CShen\, defined as 

\begin{equation}
\label{eqn:Cshen}
\CShen \equiv \Reff / \RShen,
\end{equation}
	
\noindent where \RShen\ is the radius-mass relation for local early-type galaxies from SDSS by \citet{shen2003}. Hence, mass discrepancy grows with  compactness.  There are two implications. First, the choice of $\kpar=5$ in equation~\ref{eqn:kpardef} does not provide good total dynamical masses. Second, \kpar\ is very unlikely to be constant, which implies non-homology. Overall, a clear understanding of the applicability of the widely-used equation~\ref{eqn:kpardef} (with a constant value of \kpar) is lacking. 

The goal of the present work is to explore from a computational standpoint how the structure of massive galaxies influences the dry merger process and evolves with it. Our study is not carried out as a full statistical analysis to compare with observational samples, but as an estimation of the effect and evolution of two galaxy parameters in a few realistic merger histories. \newline
The first such parameter is the S\'ersic index \citep{sersic1968}, which characterizes the changing slope of the surface brightness profile of the galaxy with radius. By varying the initial S\'ersic index of the main galaxy in our merger trees, the dependence of the dry merging process on this parameter is evaluated. This effect has not been studied before. The evolution of the index and of the density profile in general during the simulations is also analysed. 
\newline
The second parameter employed to analyse the evolution of the galaxy in the dry merging process is \kpar\ from equation~\ref{eqn:kpardef}. 
\kpar\ depends on the S\'ersic index and the shape and size of the dark matter halo, as well as on the orbital structure, all of which can vary substantially across different galaxies. In order to allow astronomers to use the virial mass estimator more effectively, we report on the dependence of \kpar\ on the S\'ersic index and on the dark matter fraction within \Reff, for equilibrium systems with S\'ersic indices and dark-matter fractions ranging from compact to normal massive ellipticals. We also report measurements of the evolution of \kpar\ during the dry merger evolution, and explore what mass parameter is best approximated when equation~\ref{eqn:kpardef} is used with $\kpar = 5$. 

Here follows the structure of the paper. 
Section~\ref{sec:virialeqns} presents the equations for the virial mass equations. 
Section~\ref{sec:simulations} presents the details of the simulations used for this study. 
In Section~\ref{sec:results} the main results of the simulations are presented and discussed. 
Section~\ref{sec:kpar} analyses the evolution of homology during the simulations, as characterized by $K$, and presents relations for the dependence of \kpar\ on homology-invariant properties of the galaxy. 
Shortcomings are discussed in Section~\ref{sec:shortcomings}, while
Section~\ref{sec:conclusions} sums up the conclusions. In Appendix A, the code which generated the galaxy models in this study is presented. 
\newline

\section[]{the virial scaling of mass, radius and velocity dispersion}
\label{sec:virialeqns}

Eqn.~\ref{eqn:kpardef} derives from the balance of potential and kinetic energies in steady-state gravitational systems, encapsulated in the virial theorem \citep{bt1987},

\begin{equation}
\label{eqn:virial}
\EPOT + 2\,\EKIN = 0,  
\end{equation}

\noindent when total potential and total kinetic energies are expressed in terms of observables, such that:

\begin{equation}
\label{eqn:ekin}
\EKIN \equiv \displaystyle\sum_{i} m_i\, v_i^2 /2 = \frac{1}{2} \Mtot \langle v^2\rangle = \frac{1}{2}\, \kkin\, \Mtot\, \sigmae^2
\end{equation}

\noindent and 

\begin{equation}
\label{eqn:epot}
\EPOT \equiv \displaystyle\sum_{i,j} G m_i m_j / r_{ij} = -  \frac{G\, {\Mtot}^2}{\rgrav} = - \frac{1}{\kstr}  \frac{G {\Mtot}^2}{\Reff}.
\end{equation}
In the equations above, the total mass \Mtot\ includes luminous and dark matter, $\langle v^2\rangle$ is the mean square speed of the mass elements, and \rgrav\ is the gravitational radius, defined by equation~(\ref{eqn:epot}). The indices $i$ and $j$ represent the single elements (particles) of the gravitational system. Equations~(\ref{eqn:ekin}) and (\ref{eqn:epot}) define \kkin\ and \kstr\ as $\kkin \equiv \langle v^2\rangle / \sigmae$, and $\kstr \equiv \rgrav / \Reff$. 
Substituting into equation~(\ref{eqn:virial}) and defining:

\begin{equation}
\kpar\ \equiv \kstr \kkin
\end{equation}
we obtain the definition of \kpar\ (Eq. \ref{eqn:kpardef}) again. \kpar\ thus does not depend on the size or mass of a galaxy, but only on the internal distribution of luminous and dark matter and on the internal orbital structure. Since \kpar\ encapsulates the scaling of observables \Reff\ and \sigmae\ to the internal mass and velocity distribution parameters \rgrav\ and $\langle v^2\rangle$, 
different values of \kpar\ imply non-homology\footnote{Since \kpar\ is the product of two terms (structural and kinetic), the converse statement, namely that equal values of \kpar\ imply homology, is not generally true.}.  

Due to its scaling invariance, \kpar\ and Equation \ref{eqn:kpardef} have also been widely used to provide dynamical-mass estimates in terms of the observables \Reff\ and \sigmae.
\citet{cappellari2006,atlasxv} provide one of the most popular calibrations of \kpar. 
Working with 2D spectroscopy of local-universe early-type galaxies from the ATLAS3D survey, they derive
mass-to-light ratios within a sphere of radius \Reff, $(M/L)_{r<\Reff}$, and thus dynamical mass estimates, using anisotropic Jeans modelling of their kinematics. These mass-to-light ratios are found to scale with \Reff\ and \sigmae\ as:
\begin{equation}
(M/L)_{r<\Reff} \approx 5.0 \frac{\Reff\,\sigmae^2}{G\,L},
\end{equation}

\noindent where $L$ is the total luminosity in the $I$ band calculated through the Multi-Gaussian Expansion technique \citep{mge}. \newline Since

\begin{equation}
\label{eqn:masstolightcomparison}
(M/L)_{r<\Reff} \approx (M/L)_{r<\rhalflight}, 
\end{equation}

\noindent where \rhalflight\ is the radius of the volume enclosing half the total light, it follows that 

\begin{equation}
\Mhalfdyn \approx 2.5 \frac{\Reff\,\sigmae^2}{G}
\end{equation}

\noindent and, given that in massive early-types the luminous component dominates the density in the central parts, the total stellar mass of the galaxy is expected to be

\begin{equation}
\label{eqn:mlumvirial}
\Mlum \approx 5.0 \frac{\Reff\,\sigmae^2}{G}.
\end{equation}

\citet{atlasxv} use the mass-to-light ratios from their Jeans anisotropic modelling to define a mass parameter named \Mjam, such that 

\begin{equation}
\Mjam \equiv (M/L)_{r<\Reff} \times L
\label{eqn:Mjamdefinition}
\end{equation}

Given the above equations, \Mjam\ is approximately 

\begin{equation}
\label{eqn:mjamvirial}
\Mjam \approx 5.0 \frac{\Reff\,\sigmae^2}{G}, 
\end{equation}

\noindent and approximately describes either the total luminous mass, or twice the dynamical mass within \rhalflight:

\begin{equation}
\label{eqn:MjamMlum}
\Mjam \simeq 2\,\Mhalfdyn \simeq \Mlum
\end{equation}

\noindent Because of their simplicity, one would like to use equations~(\ref{eqn:mlumvirial}) and (\ref{eqn:mjamvirial}) to obtain mass measurements for galaxies at high redshift. What is the accuracy of these scalings, which are derived on local massive early-type galaxies, for galaxies at high redshift? Our models can be used to provide three answers: first, to quantify how the virial scaling coefficient \kpar\ of spherical galaxy+halo models varies with galaxy parameters such as S\'ersic index and dark matter fraction. Second, to quantify the evolution of \kpar\ along a merger sequence. And, third, to see how accurately the \Mjam\ parameter measures \Mlum\ or \Mhalfdyn\ , i.e., to measure the errors introduced in the assumption in equation~(\ref{eqn:masstolightcomparison}). We address these three points in subsections \ref{sec:kn}, \ref{sec:klummergerevol} and \ref{sec:MjamVsMlumMdyn} below. We use the above results to show in subsection~\ref{sec:meaningof5sigma2Re} what mass parameters are best approximated by the expression $5\,\Reff\,\sigma^2\,/\,G$.

\section[]{Simulating the growth of compact ETGs}
\label{sec:simulations}

\subsection{Initial galaxy models}
Our initial galaxy models for the merger simulations are two-component $N$-body spherical non-rotating models. A full description is given in Appendix A, so only a brief summary is presented here. The luminous component follows a \citet{ps1997} density profile:

\begin{equation}
\rho_\mathrm{lum}(r) = \rho_\mathrm{lum}(0) \left({b_n}^n \frac{r}{\Reff}\right)^{-p} \, \exp\left\{-b_n\,\left(\frac{r}{\Reff}\right)^{1/n}\right\}
\label{eqn:prugnielsimien}
\end{equation}

\noindent which, in projection, accurately follows the \citet{sersic1968} surface brightness profile

\begin{equation}
\mu(R) =  \mu(0)\,\exp\left\{-b_n\,\left(\frac{R}{\Reff}\right)^{1/n}\right\}
\label{eqn:prugniel}
\end{equation}

\noindent where \Reff\ is the radius enclosing half the total mass in projection and \nsers\ is the S\'ersic index which drives the central concentration of the profile; see Appendix A for the definitions of the other parameters in both expressions above, and \citet{trujillo2002} for an alternative deprojection of the S\'ersic profile whose higher-accuracy is however not suitable for the integrations needed to compute the model's distribution function. 

The second component represents dark matter. Its particles follow the \citet{hernquist1990} profile

\begin{equation}
\rho_\mathrm{dark}(r) = \frac{M}{2\pi {\aH}^3} \, \left(\frac{r}{\aH}\right)^{-1} \left( 1 + \frac{r}{\aH} \right)^{-3}
\label{eqn:hernquist}
\end{equation}

\noindent where $M$ is the total mass of the dark matter halo and \aH\ is roughly the radius where the profile slope steepens from -1 to -4, and is related to the halo half-mass radius as $\aH = \rhalflight / (1 + \sqrt{2})$. The Hernquist profile was chosen over the NFW profile \citep{nfw} because, unlike the latter, it has a finite total mass, and is thus more convenient for constructing $N$-body models. It is also equivalent to a NFW profile within its $r_{1/2}$. 

The models are generated in a state of near equilibrium, using a variant of the Osipkov-Merrit model \citep{bt1987}. The distribution function of each component is calculated separately with the Eddington formula:
\begin{equation} f({\cal E}) = \frac{1}{2 \sqrt{2} \, \pi^2} \int_0^{\cal E} \frac{d^2\rho}{d\Psi^2} \, \frac{1}{\sqrt{\Psi-{\cal E}}} \, d\Psi, \end{equation}
where in the second derivative $\rho$ is the density profile of the single component and $\Psi$ is the gravitational potential of the whole system. The velocities of the particles are then assigned according to this distribution function, so that the system is in near equilibrium. More details, including the options for radial and tangential anisotropy, can be found in Appendix A. 
In summary, a model is characterized by 7 parameters: the masses of the luminous and dark components (\Mstars, \Mdark), their size parameters (\Reff, $a$), their anisotropy radii, and the S\'ersic index \nsers\ of the luminous component. The generated models have excellent stability properties. All the models employed in this paper were isotropic. 

\subsection{Physical properties of the initial galaxies}
Following the recipe of \citet{tapia2013}, the mass of the galaxies and the spatial configuration of the mergers have been obtained identifying two merger trees in the GALFOBS project (Galaxy Formation at Different Epochs and in Different Environments: Comparison with Observational Data), a series of N-body+SPH cosmological simulations with $(2^{9})^3$ baryonic particles and $(2^{9})^3$ dark matter particles. The two merger trees will be identified as A and B, and correspond to the merger trees 7 and 3 of \citet{tapia2013}, respectively. The size of the progenitor galaxies was chosen to emulate the observed compact elliptical galaxies. The size of the satellites was instead taken from the empirical law by \citet{shen2003}, which gives us the effective radius of elliptical galaxies as a function of mass:
\begin{equation} \frac{\Reff}{\rm kpc} = 1.15 \left( \frac{\Mlum}{10^{10} {\rm M}_{\odot}}\right)^{0.56} . \end{equation}
This formula was obtained by fitting data of early-type elliptical galaxies in the Sloan Digital Sky Survey. It was derived from data of the local universe, therefore it needs to be rescaled taking into account the evolution with redshift. To this purpose, we use the relation by \citet{trujillo2006}:
\begin{equation}\Reff (z) = \Reff(0) \cdot (1+z)^{-0.45\pm0.10} . \end{equation}
These relations are not the most recent in the literature, but they are sufficient for the scope of this work. The satellites have been generated with a S\'ersic index of $n=2$, which is a realistic value forintermediate mass elliptical galaxies \citep{caon1993}. 
\newline
All the initial models are isotropic. Before the beginning of the main simulations they have also been left evolving on their own for $\sim 170 \, \rm Myr$, to ensure stationarity. All of their properties are shown in Table \ref{tab:tapia1} and Table \ref{tab:tapia2}.
\begin {table}
 \caption {\small Input parameters of the initial galaxies.} \label{tab:tapia1} 
\begin{center}
  \begin{tabular}{ | c | c | c | c | c | c | c |}
    \hline
     ID & $\Mlum $ & $\Reff $ & $r_{1/2, {\rm dark}} $ & $ n $ & z\\ 
     & $(10^{11} {\rm M}_{\odot})$ & $({\rm kpc})$ & $({\rm kpc})$ &  &  \\ \hline \hline
    Progenitor A & 1.68 & 1.020 & 66.6 & 1,2,4 & 2.5\\ \hline
    Satellite A1 & 0.480 & 1.860 & 58.1 &  2 & 1.30\\ \hline
    Satellite A2 & 0.166 & 1.094 & 42.6 & 2 & 1.00  \\ \hline
    Satellite A3 & 1.661 & 5.189 & 124.3 & 2 & 0.10 \\ \hline
    Progenitor B & 0.520 & 0.960 & 44.7 & 1,2,4 & 2.5\\ \hline
    Satellite B1 & 0.371 & 1.928 & 64.9 & 2 & 0.55 \\ \hline
    Satellite B2 & 0.162 & 1.263 & 49.2 & 2 & 0.40 \\ \hline
    Satellite B3 & 0.527 & 2.625 & 79.6 & 2 & 0.20 \\ 
    \hline
  \end{tabular}
\end{center}
\end{table}

\begin {table}
 \caption {\small Configuration of the mergers.} \label{tab:tapia2}  
\begin{center}
  \begin{tabular}{ | c | c | c | c | c | c | c |}
    \hline
     ID &  Mass &  $R_{\rm in}$  &  $\phi$  &  $\theta$ & $v_{\rm R}$ & $v_{\rm T}$\\ 
     & ratio & (${\rm kpc}$) & (deg) & (deg) & (${\rm km \, s}^{-1}$) & (${\rm km \, s}^{-1}$) \\ \hline \hline
    \multicolumn{7}{|c|}{Merger Tree A} \\ \hline
    A1 & 3.5  & 44.93 & 0 & 0 & -322.45 & 163.43 \\ \hline
    A2 & 13.0  & 33.92 & 56 & 45 & -361.40 & 177.11 \\ \hline
    A3 & 1.4  & 57.60 & 59 & 156 & -204.46 & 179.66 \\ \hline \hline
    \multicolumn{7}{|c|}{Merger Tree B} \\ \hline
    B1 & 1.4  & 59.90 & 0 & 0 & -261.49 & 156.71\\ \hline
    B2 & 5.5  & 28.29 & 45.31 & 25 & -265.20 & 170.85\\ \hline
    B3 & 2.0  & 34.43 & 55.04 & 128 & -226.48 & 207.24\\ \hline
  \end{tabular}
\end{center}
\end{table}

\subsection{Properties of the dark matter haloes} \label{sec:dmhalos}
All the galaxy models used in our simulations have a dark matter halo whose density profile follows the Hernquist law (Equation \ref{eqn:hernquist}), with mass 10 times the mass of the luminous component of the galaxy. This fixed value, while not necessairly cosmologically accurate, results in realistic central dark matter fractions of the final galaxies. \newline
To choose the size of the halo several studies that evaluate the relation between the concentration $c$ of a dark matter halo and its mass and redshift have been considered (see \citet{coe2010} for a review). The concentration parameter is defined as $c_{200} = r_{200} / r_{\rm NFW}$, where $r_{\rm NFW}$ is the characteristic radius of the NFW profile which best fits the halo and $r_{200}$ is the radius at which the average density inside it is 200 times the critical density $\rho_c$ of the universe (i.e., the virial radius):
\begin{equation} r_{200} = \left(\frac{M_{200}}{4/3 \, \pi \, 200 \, \rho_{c}} \right)^{1/3} . \end{equation} 
This allows to convert an equation for the concentration into an equation for a size parameter of the halo, $r_{\rm NFW}$. Most of the published $c=c(M,z)$ relations are in the format $c=(A/(1+z)^B) \, M^C$ (e.g., \citet{bullock2001}). For the models in these simulations we used the one by \citet{duffy2008}, obtained through a cosmological simulation, which is the most recent in this format:
\begin{equation} c_{200} = \frac{6.71}{(1+z)^{0.44}} \left(\frac{M_{\rm halo}}{2 \cdot 10^{12} \, h^{-1} \, {\rm M}_{\odot}} \right)^{-0.091} . \end{equation}
More recent papers (e.g., \citet{prada2012}) propose a much more complicated model to determine halo concentrations, but in the range of mass covered by the galaxies in these simulations the results are very similar. The value of the critical density is a function of redshift, and has been calculated for a standard $\Lambda$-CDM cosmology with $H_0 = 70.4 \, {\rm km \, s^{-1} \, Mpc^{-1}} $, $\Omega_{\rm d} = 0.728$, $\Omega_{\rm m} = 0.272$, $\Omega_{\rm rad} \simeq 0$, and $\Omega = 1$ (Jarosik et al. 2011):
\begin{equation}\rho_{c} = \frac{3 H_0^2 }{ 8 \pi G } \, (\Omega_{\rm d}+\Omega_{\rm m} \, a^{-3}+\Omega_{\rm rad} \, a^{-4}+(\Omega-1) \, a^{-2}) ,\end{equation} 
where $a=1/(1+z)$ is the scale parameter of the universe. In order to obtain the Hernquist size parameter \aH\ from $r_{\rm NFW}$ we derived the Hernquist model with the same mass inside its Hernquist radius as the NFW model up to that same radius. It is:
\begin{equation}  \aH  =  \left(\frac{1}{\sqrt{k}} - 1 \right) \, r_{\rm NFW}, \end{equation}
where
\[ k \equiv \frac{\log 2 - \frac{1}{2}}{\log(1+c_{200})-\frac{c_{200}}{c_{200}+1}} . \]
Putting all these equations together we obtain a relation between the Hernquist parameter of a halo and its mass and redshift, which is what has been used to generate the models. The choice of using a Hernquist profile rather than an NFW for generating the dark matter haloes has minimal effects on the merger evolution, as the density of the haloes is equivalent in the region where dark matter dominates. Due to the kind of matching we applied, the Hernquist halo can be up to $40\%$ more massive than the equivalent NFW within the Hernquist radius, which impacts the value of \kpar\ of the initial models. The effect is however small ($\sim 2 \%$ for \klum\ ), and arguably a more concentrated halo within the central region is more realistic anyway, due to the compactness of the galaxy it hosts.\newline
It is worth noting that the size of the dark haloes used in these simulations depends only on their mass and redshift, not on the size of the luminous component. This means that the compact primary galaxies have a very large dark matter halo relatively to their size. This is realistic if we imagine that these galaxies have shrunk to their current size by gas dissipation. For comparison, in the simulations of \citet{tapia2013} all the galaxies were assigned a dark matter halo 4.8 times larger in half-mass radius and 10 times larger in mass than the luminous component. This difference is expected to cause lower velocity dispersions in our galaxies, as well as a smaller growth of the velocity dispersion itself during the mergers.

\subsection{Computational details} \label{sec:compdetails}
The simulations were carried out using the Gadget2 tree-SPH code \citep{gadget2} on the LaPalma cluster and on machines of the Instituto de Astrofisica de Canarias. In each merger tree the mass of the particles is set so that the primary galaxies have 50000 luminous matter particles at the beginning of the simulations: $3.35 \cdot 10^6 \,  {\rm M}_{\odot}$ in Merger Tree A and $1.04 \cdot 10^6 \, {\rm M}_{\odot}$ in Merger Tree B. The mass of dark matter particles is set as twice the one of luminous particles. The total number of particles in each Merger Tree is about 750000, of which about 200000 of luminous matter. The softening length was set at 16 pc for visible matter and at 32 pc for dark matter. All the models have been left evolving in isolation before the mergers, to make sure that they are stable. Each of the simulated merger events has been carried out indipendently, waiting for the galaxies to be fully merged before beginning the next merger. After each merger is completed, particles at more than 80 kpc from the galaxy centre are removed, and the remnant is positioned according to the initial orbital parameters of the next merger from the cosmological simulation. Almost all of the mergers were over after 6000 internal time units (about 2 Gyr). The idle times between mergers have not been simulated. The unit conversion between the internal units used in the simulations and the physical units shown in the paper is the following: [L]=$0.8 \, {\rm kpc}$, [M]=$10^{12} \,{\rm M}_{\odot}$, [T]=$0.34 \, {\rm Myr}$, [V] = $2318.14 \, {\rm km \, s}^{-1}$. Throughout the simulations energy is conserved within 0.1 per cent. \newline 

\subsection{Analysis methods}

The measurements of effective radius \Reff\ and effective velocity dispersion \sigmae\ have been carried out with a routine that mimicks the limits of real astronomical measurements due to the background luminosity. We iteratively cut off the galaxy where the circularly-averaged surface density profile gets 6 magnitudes below the effective surface density, and proceed to measure the effective radius from the curve of growth of this cut model. We then calculate \sigmae\ by interpolating the integrated line of sight velocity dispersion profile with a polynomial and evaluating it at \Reff.  The reported values of \Mlum\ , \Reff\ and \sigmae\ and their errors are the averages and dispersions, respectively, over measurements for 30 random orientations of the models. 
The stellar mass \Mlum\ is the luminous mass inside the cutoff radius. The S\'ersic index \nsers\ was determined by fitting the S\'ersic law to the circularly-averaged surface mass density profile of the galaxies, using the Levenberg-Marquardt non-linear fitting method \citep{numrec}. The S\'ersic index, the effective radius and the effective luminosity were used as free parameters, and all the fits were carried out in the radial range 0.08 kpc - 80 kpc. The displayed value of \nsers\ is again the average over viewing angles. In order to evaluate how well the S\'ersic laws represent our surface density profiles we computed the reduced $\chi^2$ value for every fit, using the formula:
\begin{equation}
\chi_{\nu}^2 = \frac{1}{N_{\rm data}} \sum_{\rm data} \frac{(\Sigma_{\rm data}-\Sigma_{\rm fit} (R_{\rm data}) )^2}{\sigma_\Sigma^2}
\end{equation}
The $\chi_{\nu}^2$ values displayed in Table \ref{tab:results1} are the average over different viewing angles. The best-fitting values of \Reff\ from the S\'ersic fits are generally compatible with the ones from the curve of growth, which are shown in the tables. They can however differ significantly in the cases where the galaxies are not well fitted by a S\'ersic law. The errors displayed in the tables and figures of this paper have been calculated by summing quadratically the dispersion of the quantity over the different viewing angles with the intrinsic measurement error (averaged over the viewing angles). It should not come as a surprise that these errors are many times smaller than typically found on observational data: masses and velocity dispersions are very well-defined quantities in $N$-body models, and are not subject to uncertainties of measurement on real systems.

\begin{table*}
\caption {\small Physical properties (stellar mass, effective radius, S\'ersic index and velocity dispersion) of the main galaxy after each stage of each merger tree. 0=progenitor galaxy, 1=remnant of the first merger, 2=remnant of the second merger, 3=final remnant, 3p= material from the progenitor in the final remnant. } \label{tab:results1} 
\input{table3}
\end{table*}

\begin{figure}
\begin{center}
\includegraphics[width=0.5\textwidth]{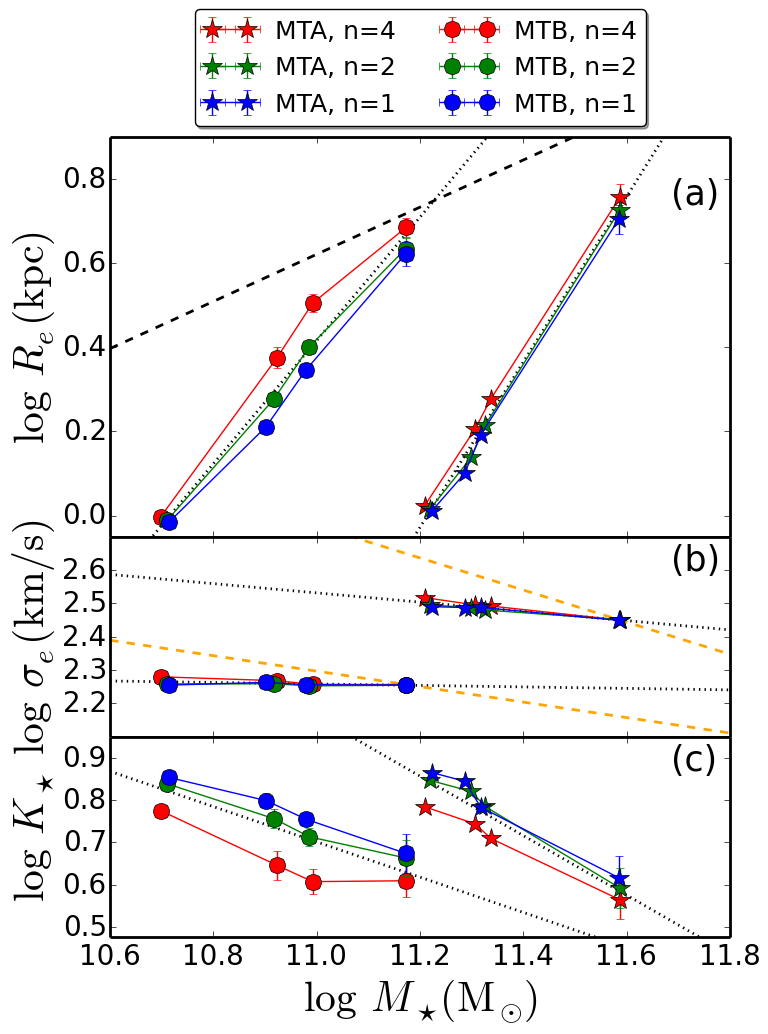}
\caption{\small (a) Luminous mass vs. Effective radius for the six merger trees, color-coded as in the legend. Dotted lines show power-law fits $\Reff \propto \Mlum^\alpha$ from orthogonal regressions to data from each tree. The dashed line traces the mean mass-radius relationship for SDSS early-type galaxies from \citet{shen2003}. (b) Luminous mass vs. effective velocity dispersion $\sigmae$ for the six merger trees. The dotted lines trace power-law fits from orthogonal regressions. The dashed lines trace the locus of velocity dispersion that would be expected from homology. 
(c) Luminous mass vs. the virial constant $\klum \equiv G\,\Mlum/ \Reff\,\sigmae^2$ (eqn.~\ref{eqn:computeklum}) for the six merger trees. The dotted lines are power-law fits from orthogonal regressions.}
\label{fig:massreffsigma}
\end{center}
\end{figure}

\begin{figure*}
\centering
\includegraphics[scale=0.6]{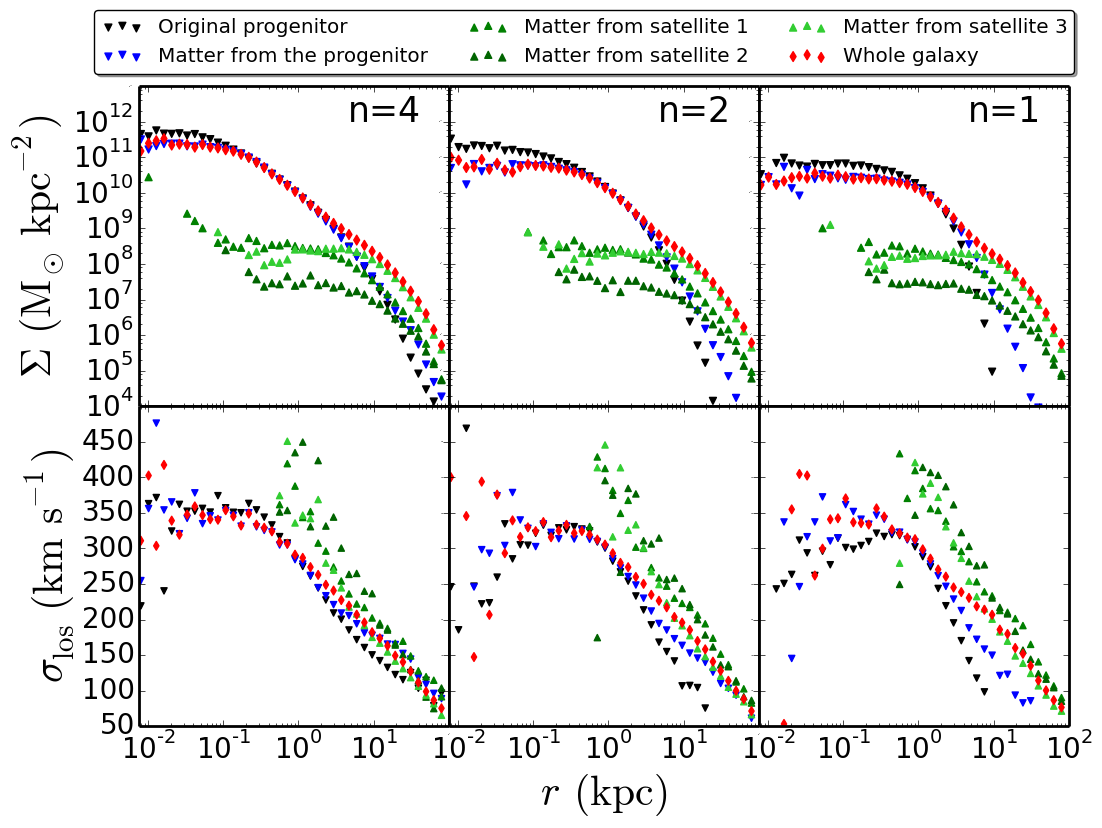}
\caption{\small Upper panels: Surface mass density profile of the final galaxy of Merger Tree A in the three cases ($n=1,2,4$) compared to the contributions from the progenitor and each satellite. The surface brightness of the original progenitor (before the mergers) is also shown for comparison. Lower panels: line of sight velocity dispersion profile of the final galaxy and its components taken separately. The satellite data
inside 0.5 kpc are omitted given the low density of satellite particles at small radii.}
\label{densvdispres1}
\end{figure*}

\begin{figure*}
\centering
\includegraphics[scale=0.6]{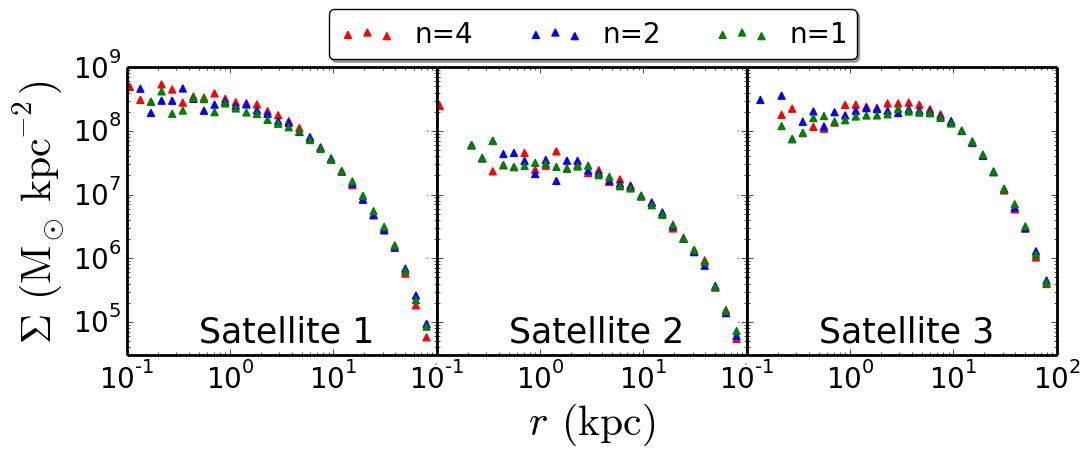}
\caption{\small 
Comparison of the contributions of each satellite to the surface mass density profile of the final galaxy of Merger Tree A in the three cases ($n=1,2,4$). }
\label{satcomp}
\end{figure*}

\section[]{Results}
\label{sec:results}

Results are summarized in Table \ref{tab:results1}, which lists luminous mass, effective radius, velocity dispersion and S\'ersic index for each merger tree. Values are given for the initial model,  after each merger is complete, and at $z=0$. For comparison, Table \ref{tab:results1} also shows the measurements of \citet{tapia2013} for the initial and final systems. 
Our analysis focuses less on the evolution with redshift, which has been amply documented in previous works with more statistics \citep[e.g.,][]{oser2012,tapia2013}, and more on the relations between mass, size, velocity dispersion and S\'ersic index, which provide clues on the types of equilibria that are generated by the merger processes modelled here. 

\subsection{Mass and size growth}
\label{sec:sizegrowth}

The evolution of \Reff\ with \Mlum\ is depicted in Figure~\ref{fig:massreffsigma}(a). 
\Reff\ gets significantly larger in the six merger trees. Each model starts in the region of compact massive galaxies, $\Reff = 1$ kpc, and finishes near the  mass-size relation for early-type galaxies in the local universe, outlined in the figure with the \citet{shen2003} relation. The average ratio of the size increase is about 4.5, while the average ratio of the mass increase is 2.5. The total growth is similar to that found by other authors, although it is somewhat smaller than what \citet{tapia2013} found for the same merger trees, possibly because of the different initial models. The growth closely follows a power-law of the luminous mass, $\Reff \propto (\Mlum)^\rho$; orthogonal fits yield slopes $\rho$ from 1.38 to 1.97 for the six merger trees, with a mean of $\rho = 1.68$.  
Already noted by many authors, the pronounced growth of \Reff\ is the result of the deposition of accreted matter in the outer parts. For illustration, Figure \ref{densvdispres1} shows the surface mass density and velocity dispersion profiles of the final galaxy of Merger Tree A for the three different initial S\'ersic indices. The profile of the final galaxy is compared with the profiles of the matter from the progenitor and from the satellites within the final galaxy, as well as with the profile of the progenitor before the mergers took place. As can be seen in the figure, the matter from the satellites dominates in the outer parts of the final galaxy, and the core of the progenitor is largely unaffected. This shows that size growth is mostly determined by the accretion of matter from the satellites in the outer parts of the primary galaxy. 

Table~\ref{tab:results1} shows that the growth process has a very small dependency on the S\'ersic index of the initial galaxy. Galaxies with larger initial \nsers\ grow slightly larger than galaxies with smaller initial \nsers. The dependence on the particulars of the merger tree is however much more important: the average growth rate for the two merger trees taken separately is $\rho_\mathrm{A} = 1.96 \pm 0.06$ for tree A and $\rho_\mathrm{B} = 1.46 \pm 0.09$ for tree B. The very limited statistics we have already show that growth-rate differences due to the progenitor S\'ersic index are much smaller than those arising from the merger history. 
To illustrate why the dependency of the growth rate on the primary S\'ersic index is so small, we plot in Figure~\ref{satcomp} the surface density distribution of matter originally belonging to each of the three satellites in merger tree A. In each of the top panels we compare the surface densities after mergers with primaries with $n = 1, 2, 4$, while the lower panels show the corresponding line-of-sight velocity dispersion profiles. Clearly, surface density and velocity dispersion profiles vary significantly from satellite to satellite, as would be expected from differences in mass ratio and merger orbits. But for each satellite, 
 the surface densities and velocity dispersion profiles of the accreted matter do not change significantly with the S\'ersic index of the primary. 
The reason for this behaviour is likely to be that most of the orbital energy of the satellite galaxies is absorbed by the dark haloes rather than by the luminous components, given the small size of the compact primaries.

\subsection{Evolution of the velocity dispersion and the virial coefficient}
\label{sec:sigmagrowth}

Figure~\ref{fig:massreffsigma}(b) shows the evolution of the effective velocity dispersion \sigmae\ with \Mlum. We find a slight \sigmae\ decrease as mass grows, compatible with the decrease of \sigmae\ with redshift measured by \citet{oser2012}, and contrary to the increase of \sigmae\ with mass found by \citet{nipoti2009} and \citet{tapia2013}. The difference between these studies is probably related to the different sets of initial conditions. \citet{tapia2013} used the same stellar masses, sizes and orbital parameters of this paper, so we can assume that the increase in \sigmae\ they observe is caused by their choice of more compact satellites (built according to a Jaffe profile) and/or by the higher concentration of their dark matter haloes. More compact satellites would plunge deeper in the potential of the main galaxy, while more compact dark matter haloes mean more mass in the outer parts of the galaxy; both of these effects contribute to an increase in \sigmae. Our velocity dispersion trend is well modelled with a power-law $\sigmae \propto (\Mlum)^\Sigma$, with $\Sigma_\mathrm{A} = -0.14 \pm 0.02$ and $\Sigma_\mathrm{B} = -0.02 \pm 0.01$ for merger trees A and B, respectively. 
Even though the entity of this small drop in \sigmae\ seems to depend on the S\'ersic index of the progenitor, this dependence is driven uniquely by the slightly higher \sigmae\ of initial models with higher \nsers. 
Finally, we show in Figure~\ref{fig:massreffsigma}(c)  the evolution of the virial coefficient \klum\ (eqn.~\ref{eqn:computeklum}). \klum\ closely follows a power-law $\klum \propto (\Mlum)^\kappa$, and from the \klum\ definition its exponent follows 

\begin{equation}
\kappa = 1 - \rho - 2\,\Sigma
\label{eqn:KLexponents}
\end{equation}

\noindent Clearly, \klum\ does not remain constant, a sign that merger products are not homologous with precursors. Figure~\ref{fig:massreffsigma} already shows that homology is not fulfilled, because either size growth is too pronounced for the shallow change in velocity dispersion, or velocity dispersion does not decrease as much as would be necessary for the given size growth. We plot (dashed lines in Figure~\ref{fig:massreffsigma}[b]) the trend of velocity dispersion along the merger sequence which would correspond to homology between precursors and remnants. The evolution of \klum\ along a merger sequence was also studied in \citep{nipoti2009}, where the parameter appears as $c_\star= 1 / 2 \klum\ $. They also find a decrease of \klum\, albeit weaker than in our simulations ($\kappa=-0.22$, compared to $\kappa=-0.52$ in our case), due to smaller size growth they observe. This difference likely has to do with their different set of initial conditions, specifically the fact that their satellite galaxies are as compact as the progenitors. We expand the analysis of \klum\ in Section \ref{sec:kpar}.

\subsection{Evolution of the S\'ersic index} \label{sec:sersicev}
\begin{figure}
\centering
\includegraphics[scale=0.45]{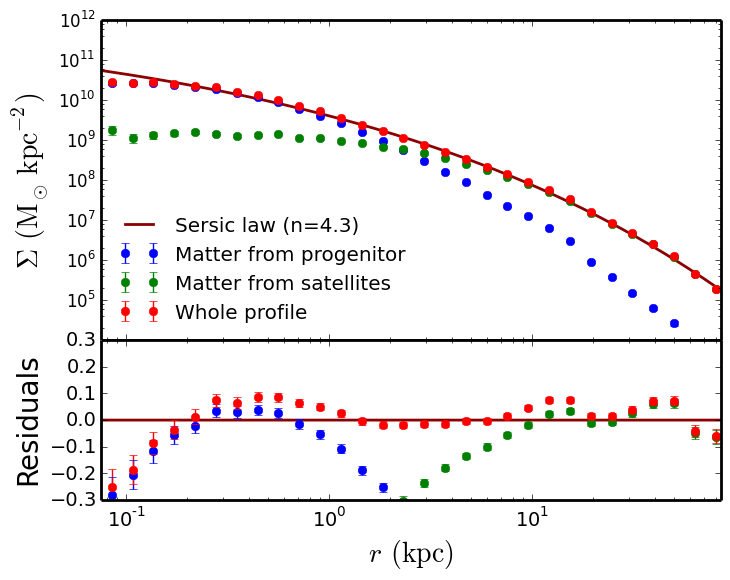}
\caption{\small Fitting with a S\'ersic law of the remnant galaxy of the Merger Tree B with initial $n=2$. The result of the fit is $n=4.3$ with ${\chi_{\nu}}^2 \sim 32.7$. }
\label{sersicfitres}
\end{figure}

The S\'ersic index of the main galaxy grows significantly in all cases. This is caused by both the matter added in the outskirts by the satellites and by the dense cores, which become `smaller' with respect to the new effective radius. Because of this composite nature, the surface brightness profile of the final galaxies is not well fitted by a S\'ersic law, but shows two `bumps' corresponding to the core and to the matter from the satellites. Figure \ref{sersicfitres} shows the profile of one of the remnant galaxies (the Merger Tree B, $n=2$ case) fitted with a S\'ersic law, and the residuals of the fit. This causes the very high ${\chi_\nu}^2$ values in Table \ref{tab:results1}. After the first mergers have occurred, all our galaxies have S\'ersic index fairly close to $n=4$, and the difference between the progenitors essentially vanishes. 

\citet{hilz2013} also studied the evolution of the S\'ersic index in dissipationless mergers, and like in our case found that it consistently increases. They also found that the entity on this increase depends on whether the mergers are major (from $n=4$ to $n=7$) or minor (from $n=4$ to $n=9.5$). Our galaxies grow through both minor and major mergers, and the extent of the increase of their S\'ersic index is not as great: in the two cases with $n=4$ progenitors, it reaches $n=7.5$ and $n=5.2$ respectively. \citet{taranu2013} measured the S\'ersic index of merger remnants of spiral galaxies with $n=1$ and $n=4$ bulges. The ellipticals they obtained had on average index $n~3$ and $n~5$ in the two cases respectively. The specific final value of the S\'ersic index likely depends on many details of the merger history, but dry mergers seem to always cause it to grow. \citet{nipoti2003} found that \nsers\ can decrease in head-on minor mergers, but increases otherwise.

\subsection{Evolution of the dark matter fraction} \label{sec:fdm}
The dark matter fraction within the effective radius, $\fdm (r < \Reff)$, consistently increases along the merger trees, from $0-5 \%$ to $20-30 \%$. The values can be found in Table \ref{tab:ksim}. This increase is mostly due to the larger growth of the luminous effective radius compared to the growth in size of the dark matter component. The expansion of the luminous body of the galaxy engulfs more and more of the dark matter, increasing the value of $\fdm (r < \Reff)$. The effect is even stronger for the dark matter fraction within the full luminous body of the galaxy, $\fdm (r < \rcutoff)$, which increases from $20-50 \%$ to about $75 \%$. Other studies \citep{ntb2009, hilz2013} also found a similar increase, even though their initial models had different (higher) dark matter fractions.

\newpage

\section{Virial masses and homology in the evolution of compact ETGs}
\label{sec:kpar}

\begin{figure}
\centering
\includegraphics[scale=0.45]{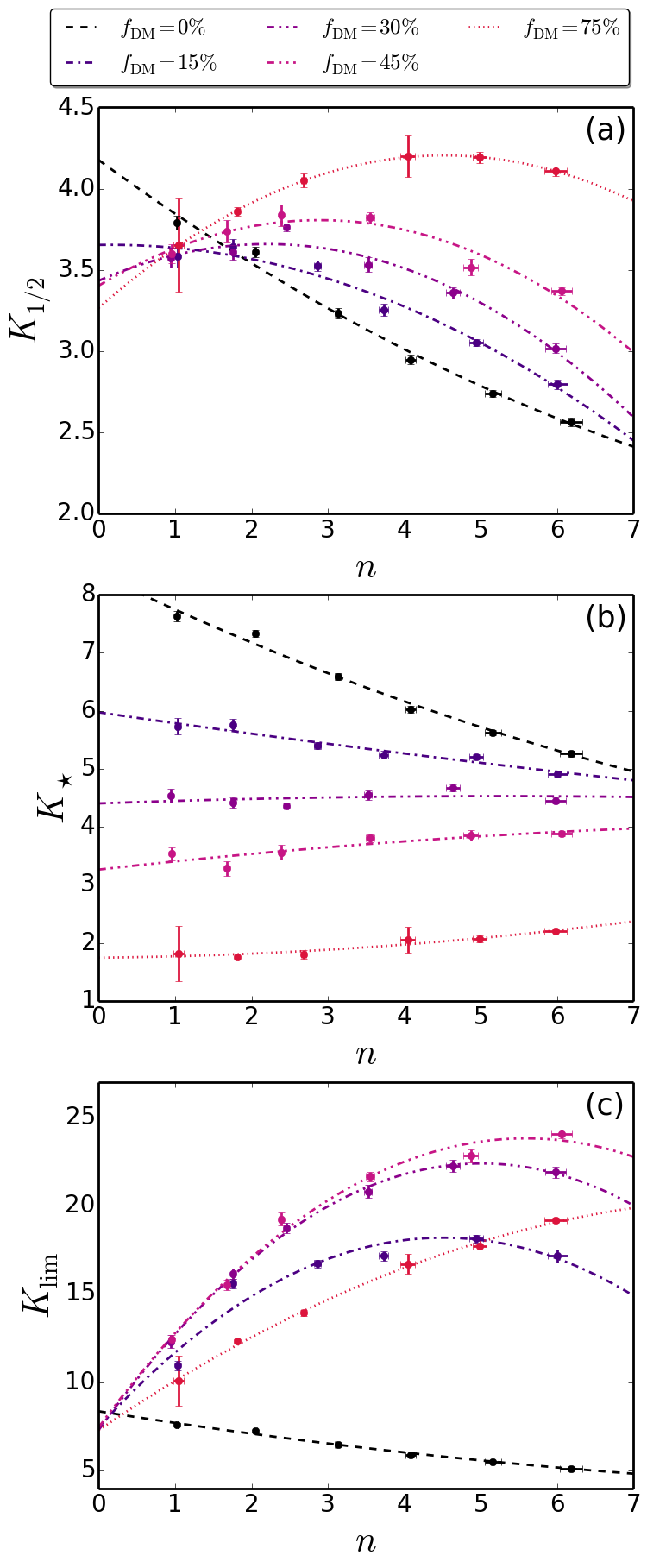}
\caption{\small \textit{(a): } Values of \khalfdyn\ obtained for dynamically stable isotropic models with S\'ersic index \nsers\ and different dark matter fractions within \Reff. Dashed lines are the best polynomial fits \klum(\nsers)  for each dark matter fraction, which are presented in Table \ref{tab:knrel}. In all models the total dark matter mass is 10 times the luminous mass of the galaxy. The size of the halo is adjusted to obtain the specific dark matter fraction within \Reff. \textit{(b): } same for \klum\ .   \textit{(c): } same for \klim.   }
\label{fig:knrelax}
\end{figure}

We now turn to the analysis of virial mass estimators $M = \kpar \Reff \sigmae^2 / G$, with the aim of providing the virial coefficient \kpar\ for a variety of useful mass parameters, and to infer limits to the validity of the homology hypothesis. 

We measure three different virial coefficients \kpar. First, \khalfdyn\ associated to \Mhalfdyn, the dynamical (luminous plus dark) mass within the 3D half-light radius \rhalflight,  
\begin{equation}
\label{eqn:computekhalf}
\khalfdyn \equiv \frac{G \, (\Mdyn)_{r < \rhalflight}}{\Reff \, \sigmae^2}.
\end{equation}

\noindent The interest of  \khalfdyn\ stems from the fact that it may be calibrated on real data, given that \Mhalfdyn\ may be inferred from kinematic measurements and dynamical modelling \citep{cappellari2006,atlasxv}. 

Second, \klum\ associated  to the total luminous mass \Mlum\ :
\begin{equation}
\label{eqn:computeklum}
\klum \equiv \frac{G \, (\Mlum)_{r < \rcutoff}}{\Reff \, \sigmae^2} ,
\end{equation}

\noindent where \rcutoff\ is the radius of the sphere containing the luminous body, chosen to be the radius where the surface density profile of the luminous matter is 6 magnitudes lower than the surface density at \Reff, i.e., traces the luminous mass of the models out to where light is detected in typical deep optical images. \klum\ and \khalfdyn\ are related by 

\begin{equation}
\label{eqn:khalfdyn2klum}
\khalfdyn = \frac{\klum}{2}\,\frac{1}{1-\fdm(\rhalflight)}
\end{equation}

\noindent Third, \klim\ associated to the dynamical mass within the luminous limit of the galaxy \Mlim\ :

\begin{equation}
\label{eqn:computeklim}
\begin{aligned}
\klim & \equiv \frac{G \, (\Mdyn)_{r < \rcutoff} }{\Reff \, \sigmae^2} \\
           & = \klum\, \frac{1}{1-\fdm(\rcutoff)}.
\end{aligned}
\end{equation}

\noindent In addition, we compute Cappellari's \Mjam:

\begin{equation} 
\label{eqn:computeMjam}
M_{\rm JAM}  = \left( \frac{\Mdyn}{\Mlum} \right)_{r<\Reff} \, (\Mlum)_{r<r_{\rm cutoff}}
\end{equation}

\noindent which we compare to \Mlum, \Mhalfdyn\ and $(\Mdyn)_{r < \rcutoff}$ in Section~\ref{sec:MjamVsMlumMdyn}. \newline
For all the \kpar\ determinations, masses are directly obtained from the models by adding particle masses; luminous particles are all assumed to have the same mass.

\begin{figure}
\centering
\includegraphics[scale=0.45]{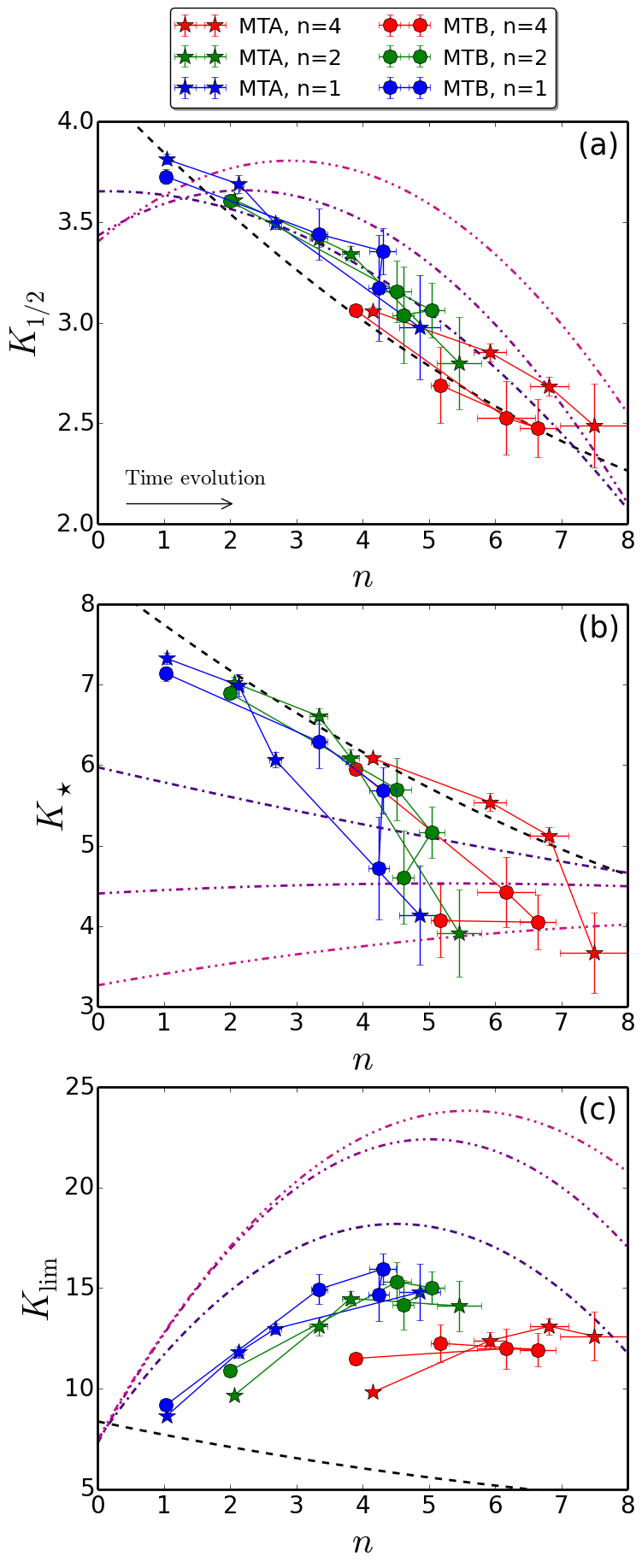}
\caption{\small \textit{(a): } Evolution of \khalfdyn\ during the merger-driven growth for all merger trees, plotted against the S\'ersic index \nsers. The general direction of evolution is from left to right. Dashed and dotted lines show the \kpar(\nsers)  relations for equilibrium S\'ersic models with typical dark-matter fractions, from Figure~\ref{fig:knrelax}. \textit{(c): } same for \klum\ . \textit{(c): } same for \klim\ .}
\label{knsimul}
\end{figure}

\begin{figure}
\centering
\includegraphics[scale=0.45]{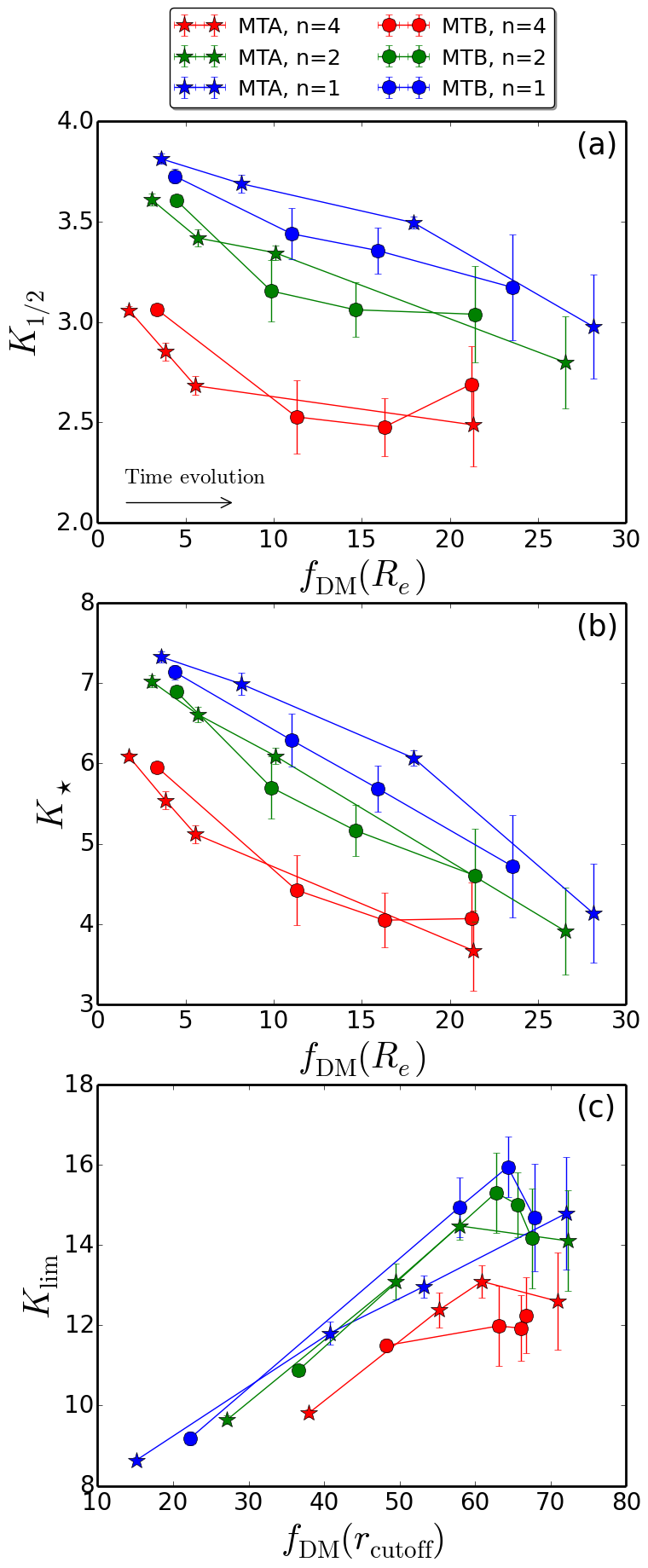}
\caption{\small \textit{(a): } Evolution of \khalfdyn\ during the merger-driven growth, plotted against the percentage dark-matter fraction within \Reff. \textit{(b): } same for \klum\ . \textit{(c): } Evolution of \klim\ during the merger-driven growth, plotted against the percentage dark-matter fraction within \rcutoff.}
\label{kDMFsimul}
\end{figure}

\begin{figure}
\centering
\includegraphics[scale=0.45]{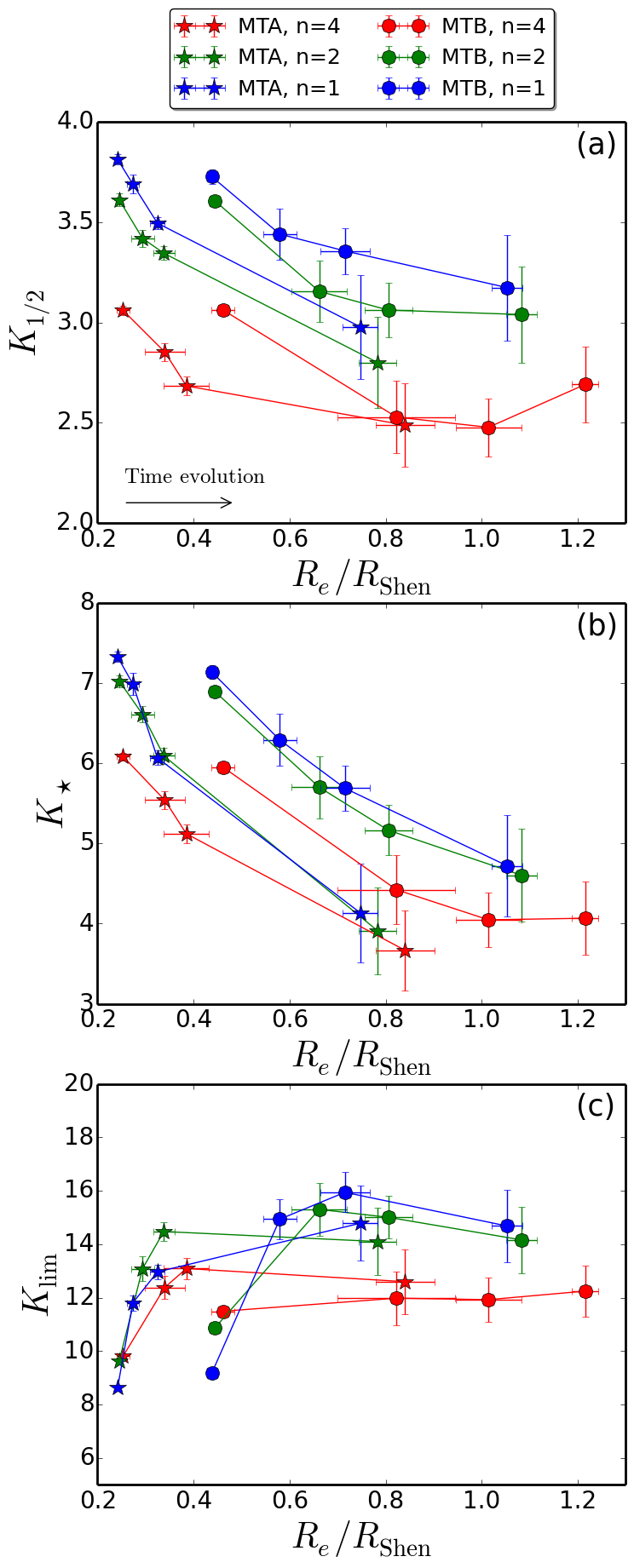}
\caption{\small \textit{(a): } Evolution of \khalfdyn\ during the merger-driven growth, plotted against the compactness index $\CShen \equiv \Reff / \RShen$. \textit{(b): } same for \klum\ . \textit{(c): } same for \klim\ .}
\label{kCshsimul}
\end{figure}

\begin{figure}
\centering
\includegraphics[scale=0.45]{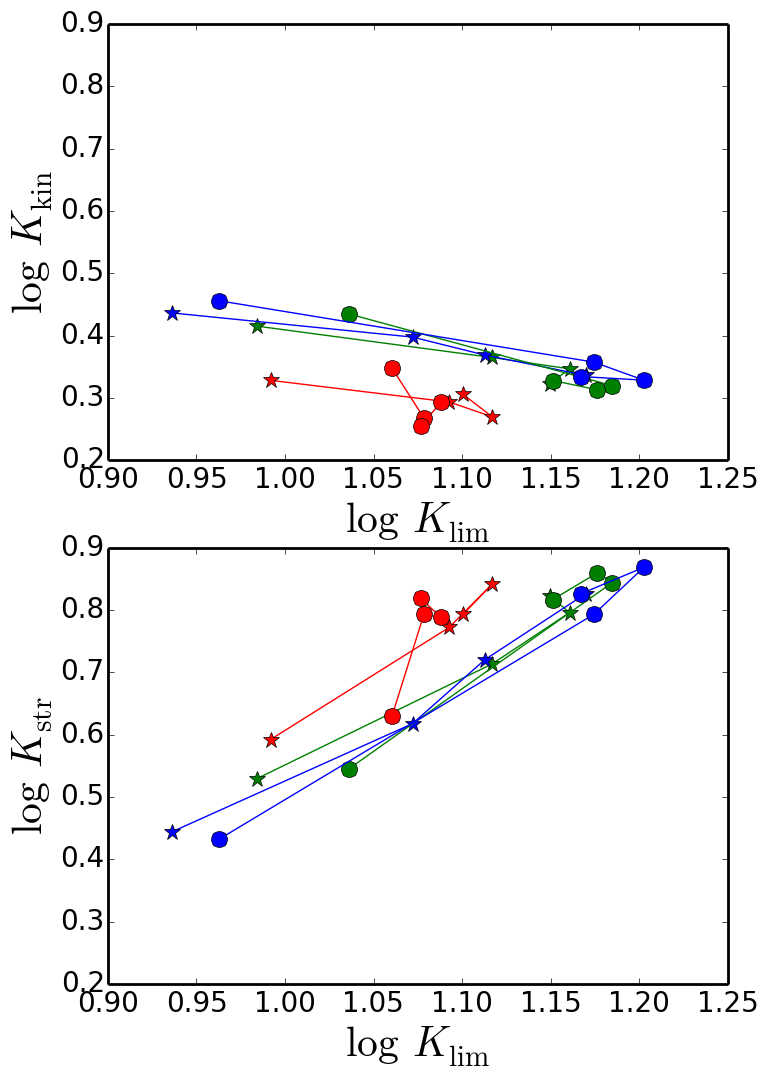}
\caption{\small Evolution of \kkin\ and \kstr\ through the dry merger simulations, as a function of \klim\ . \kstr\ is driving the evolution of \klim\ . Different markers and colors identify the different merger trees: stars are Merger Tree A, points Merger Tree B, red is initial $n=4$, green initial $n=2$, blue initial $n=1$.  }
\label{kdynstr}
\end{figure}

\begin{table*}
\caption {\small Low-order fits of \khalfdyn\, \klum\ and \klim\ to the S\'ersic index \nsers\ for equilibrium isotropic S\'ersic models inside Hernquist haloes, for a range of dark matter fractions within $\Reff$.} \label{tab:knrel}
\input{table4}
\end{table*}

\subsection{The dependence of \khalfdyn,  \klum\ and \klim\ on structural parameters} \label{sec:kn}
As noted in Section~\ref{sec:virialeqns}, the \kpar\  parameters (eqns.~\ref{eqn:computekhalf} - \ref{eqn:computeklim}) do not depend on the size or mass of a galaxy, but only  on the internal luminous and dark mass distribution and on the internal orbital structure. For an isolated isotropic galaxy model following a S\'ersic law, with no dark matter, the value of \klum\ depends only on the S\'ersic index \nsers. An empirical relation between the two can easily be obtained by measuring \klum\ using equation (\ref{eqn:computeklum}) and \nsers\ in different S\'ersic models. Such relations have already been obtained by other studies such as \citet{bertin2002}, who however used a different definition of the velocity dispersion, and \citet{cappellari2006}, who covered a different \nsers\ range than we do here. Our relation is an almost linear monotonic decreasing function, close to the one by Cappellari et al.:
\begin{equation} \label{eq:knnodm}
\klumnoDM (\nsers) =  8.34 - 0.64\, n + 0,01 \, n^2    .
\end{equation} 

\noindent \klum\ decreases with \nsers\ because a higher \nsers\ has higher inner velocity dispersion \sigmae\ for constant \Mlum\ and \Reff. We show below that equation (\ref{eq:knnodm}) provides a good approximation to the values of \klum\ for models of massive ellipticals embedded in dark matter haloes. Indeed the dark matter contribution to the gravitational potential within \Reff\ is small in massive ellipticals. Nevertheless,  predictions of \klum\ that take into account the dark matter contribution are important to account for the variation of this contribution when \Reff\ changes significantly, such as in the merger evolution of these massive, compact ellipticals. The same applies to \khalfdyn\ and \klim.

Empirically obtaining the dependence of the virial coefficients \klum, \khalfdyn\ and \klim\ on \nsers\ for a range of dark-matter fractions is straightforward, given that our model-building code generates two-component models with excellent equilibrium properties. Moreover, within the luminous body our halo density profile is a close match to a NFW profile. 
A series of isotropic models with different dark matter fractions and S\'ersic indices were generated. Different dark matter fractions \fdm(\Reff) were obtained by modifying the size of the haloes, keeping halo (and luminous) masses unmodified. The models were allowed to relax in isolation before measuring \khalfdyn, \klum\ and \klim. 

The trends of \khalfdyn, \klum\ and \klim\ with \nsers\ are shown in Figure \ref{fig:knrelax}, while polynomial fits are tabulated in Table \ref{tab:knrel}. 
Figure~\ref{fig:knrelax}(a) shows \khalfdyn. For no dark matter we obtain exactly half of the \kpar\ from equation~\ref{eq:knnodm}. As expected, at each \nsers, \khalfdyn\ increases for higher \fdm(\Reff). For \nsers\ and \fdm(\Reff) expected of massive ellipticals in the local universe, $\khalfdyn \approx 3$ yields a good approximation to \Mhalfdyn.

\klum, shown in Figure~\ref{fig:knrelax}(b), decreases with the increase of  \fdm(\Reff) for each \nsers: adding dark matter brings \sigmae\ up, an effect which is stronger at low \nsers. Since $\sigmae \propto \Mdyn (\Reff)$, this dependence roughly goes as $\klum \propto (1-\fdm(\Reff))$. With a higher dark matter fraction, \klum\ also depends less on \nsers\, and more on the mass distribution of the dark matter halo with respect to the effective radius of the luminous component.
\klum\ lies in the range $\sim$4 to 8 for dark-matter fractions expected in massive ellipticals. $\klum  \approx 5$ is obtained for high \nsers\ and $\fdm(\Reff) \leq 30$ per cent, and 
for any \nsers\ when  $\fdm(\Reff) \approx 25$ per cent. Hence, $\klum = 5.0$ yields a reasonable estimate of the luminous mass of de Vaucouleurs galaxies, but underestimates it for compact galaxies ($\fdm(\Reff) \approx 0$) when the mass distribution has low \nsers. For the latter, for the virial equation to yield \Mlum, one needs to boost \klum\ up by 50 per cent from the canonical value $\klum = 5.0$. 

The \klim\ dependence on \nsers\ is plotted in Figure~\ref{fig:knrelax}(c). Note the change in vertical scales in relation to the other panels. 
The curve corresponding to no dark matter is identical to that for \klum\ (equation~\ref{eq:knnodm}). 
For models with dark matter, because the dark matter fraction within the luminous body is much larger than that within \Reff, \klim\ is always larger than \klum. Contrary to the behaviour of \klum, \klim\ increases with \nsers\ because higher \nsers\ leads to a more extended light distribution, hence \rcutoff\ encompasses more of the dark halo mass. The extent to which \klim\ is higher than \klum\ is a measure of the amount of dark matter within the luminous body. In our merger models (Table~\ref{tab:ksim}) $\klim / \klum$ starts at $\sim$ 1.5 in the compact primaries, and evolves to $\klim / \klum \sim 3.5$ in the extended $z=0$ remnants, corresponding to final dark-matter fractions of $\sim$70 per cent. 
The values of \klim\ for our equilibrium models with dark matter, which lie in the range $\klim = 10 - 25$, imply that, if a galaxy is embedded in an extended dark matter halo,  the expression 
$M = 5\,\Reff\,\sigma^2\, /\, G$ provides an underestimate, by factors of 2 to 5, to the dynamical mass within the luminous body of the galaxy. We provide clues on the interpretation of $M = 5\,\Reff\,\sigma^2\, /\, G$ in Sec.~\ref{sec:meaningof5sigma2Re}. 

\begin{table}
 \caption {\small Evolution of the parameters \khalfdyn\, \klum\ and \klim\ measured through the dry merger simulations. The  evolution of the S\'ersic index and of the dark matter fraction within \Reff\ are also shown. M0=Initial model, M1=After merger 1, M2=After merger 2, M3=Final remnant.} \label{tab:ksim}
\input{table5}
\end{table}

\subsection{Evolution of \khalfdyn, \klum\ and \klim\ as a compact galaxy grows in size due to dry mergers}
\label{sec:klummergerevol}

We now address how much do the virial coefficients \khalfdyn, \klum\ and \klim\ evolve as a consequence of the structural and dynamical changes that occur as a galaxy grows by mergers. For the six merger trees, the virial coefficients were computed at the end of each merger and at the final configuration at $z=0$. The results are listed in Table~\ref{tab:ksim}, and plotted in Figures~\ref{knsimul}, \ref{kDMFsimul}, \ref{kCshsimul}, \ref{kdynstr} (the \klum\ values are those already plotted against \Mlum\ in Figure~\ref{fig:massreffsigma}).  In Figure~\ref{knsimul} we include the trends of each \kpar\ vs \nsers\ for equilibrium, isotropic models for various \nsers\ and \fdm(\Reff), from Figure~\ref{fig:knrelax}. In all the plots, evolution occurs from left to right. 

\subsubsection{Evolution of \khalfdyn} 
\label{sec:khalfdyn}

For \khalfdyn\ (panels [a] of Figures~\ref{knsimul}, \ref{kDMFsimul} and \ref{kCshsimul}) evolution is moderate: 10 to 15 per cent drop from start to end of each merger tree, to be compared with the drop in \klum\ (30 to 40 per cent, panels [b] of Figures~\ref{knsimul}, \ref{kDMFsimul}, \ref{kCshsimul}). One might have expected a stronger change, given that \nsers\ increases by factors of $\sim$2 and \fdm(\Reff) by factors of $\sim$8. But given the relation between \khalfdyn\ and \klum\ (Equation~\ref{eqn:khalfdyn2klum}), as the galaxy grows, the increase of \fdm(\Reff) partly compensates the drop of \klum. It is due to the double dependency of \khalfdyn\ on \Reff\ and on \fdm(\Reff) that \khalfdyn\ is the least sensitive of the three \kpar\ parameters to galaxy growth. \khalfdyn\ should therefore change very little with compactness (equation~\ref{eqn:Cshen}) and we show in Figure~\ref{kCshsimul}a that this is the case. 

\subsubsection{Evolution of \klum} 
\label{sec:klum}

The evolution of \klum\  is plotted in panels (b) of Figures~\ref{knsimul}, \ref{kDMFsimul}, \ref{kCshsimul}. \klum\ drops by a factor of 30 to 40 per cent, three times more than  \khalfdyn. At the end of the evolution, corresponding to massive, normal-size ellipticals, models show $\klum \approx 3.7-5$, not unlike the values derived by \citet{cappellari2006,atlasxv}. In initial models, corresponding to massive, compact galaxies at $z\sim 2$, \klum\ is 40 to 60 per cent higher. 

Notable in Figure~\ref{knsimul}(b) is that, as the galaxy grows by a series of mergers, \klum\ evolves away from the \klum-\nsers\ relationship for the \fdm(\Reff) of the models: at the end of the merger sequences \klum\ has dropped to values corresponding to $\fdm(\Reff) > 45$ per cent, while models have $\fdm(\Reff) \leq 25$ per cent (Table~\ref{tab:ksim}). The low values of \klum\ reflect that \sigmae\ is `too high' for the \Reff\ and \fdm(\Reff) of the post-merger galaxies. We expect this behaviour to be general for galaxies grown by mergers that deposit matter in the outer parts: the inner \sigmae\ remains `frozen' to the initial value, and therefore higher than a simple isotropic, equilibrium configuration would predict for the final, higher \Reff. 
\klum\  decreases with \fdm(\Reff) (Figure~\ref{kDMFsimul}b) as a result of the increase of \Reff\ shown in Figure~\ref{kCshsimul}b: as the galaxy accretes matter in its outer parts, more and more dark matter lies within the luminous boundary of the galaxy. 

\subsubsection{Evolution of \klim} \label{sec:klim}
\label{sec:klimmergerevol}

The evolution of \klim\ for the merger models is plotted in Figure~\ref{knsimul}c against the S\'ersic index of the luminous matter distribution. \klim\ ranges from $\sim$9 to 16, and rises during the merger evolution for models with initial $n = 1$ or 2, while it remains nearly flat for models with initial $n = 4$. Comparison with the $\klim - n$ relations (dashed lines) shows that the merger evolution leads to \klim\  values below those expected for the dark-matter fractions \fdm(\Reff) of the models.  We expect the evolution of \klim\ to be driven by the increase of the dark matter fraction within the luminous extent of the galaxy \fdm(\rcutoff), and we show in Figure~\ref{kDMFsimul}c that this is indeed the case. 

For the initial compact galaxies, $\fdm(\rcutoff) \approx 20$ (49) for $n = 1$ (4). 
The accretion of matter in the outer parts quickly expands the luminous bodies so that a larger volume of the massive dark haloes now contributes to the dynamical mass inside the luminous body (final $\fdm(\rcutoff) \approx 70$) with final $\klim \approx 12- 16$. 
While \klim\ grows in a nearly linear fashion with  $\fdm(\rcutoff)$ (Figure~\ref{kDMFsimul}c), which indicates that the total dynamical mass of the galaxy is dominated by dark matter, it shows a complex behaviour with compactness (Figure~\ref{kCshsimul}c), a sign that growth of the luminous galaxy's physical boundary is not due to an increase of \Reff\ but to the increase in the S\'ersic index \nsers.

Further information can be obtained by distinguishing the structural and kinetic components of \klim\  defined in equations~(\ref{eqn:ekin}) and (\ref{eqn:epot}). In Figure \ref{kdynstr} is plotted their evolution as a function of \klim\ for all the merger tress. While \kkin\ on average varies only by 2.9 per cent, \kstr\ varies by $\sim$25 per cent, driving the evolution of the main parameter \klim. Similar results are obtained decomposing the other \kpar\ parameters. Their evolution is thus driven by the structural change in the density profile, rather than by a change in the velocity distribution within \Reff. The final galaxies mostly keep their original kinematic properties and degree of anisotropy within \Reff.

\subsection{Is \Mjam\  a good tracer of \Mlum, \Medyn\ and \Mhalfdyn?}
\label{sec:MjamVsMlumMdyn}

\begin{figure}[h]
\begin{center}
\includegraphics[scale=0.4]{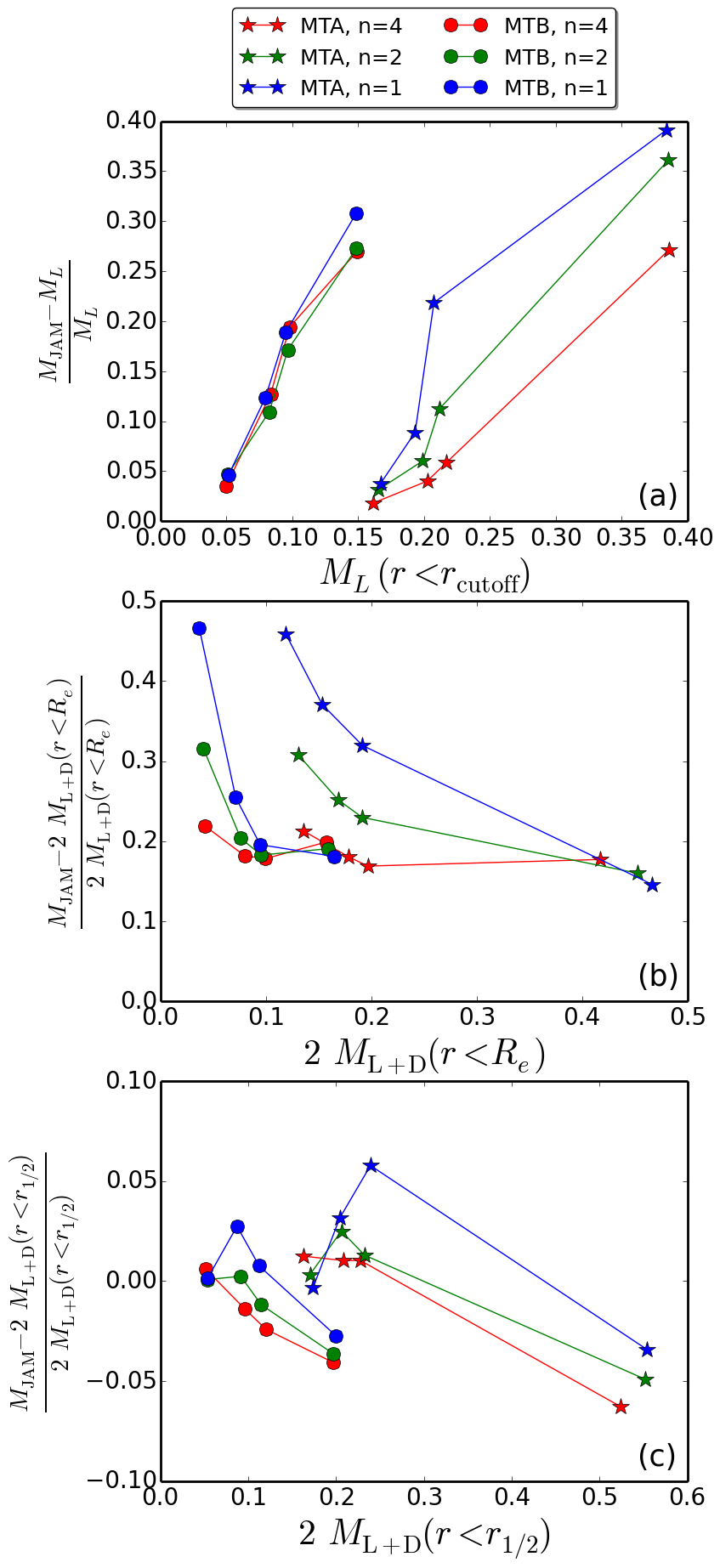}
\caption{For all the merger models, \Mjam\ is compared to: \textit{top: } the total luminous mass \Mlum; 
\textit{middle: } the dynamical mass within \Reff, \Medyn; and  
\textit{bottom: } the dynamical mass within \Reff, \Mhalfdyn. 
}
\label{fig:MjamMlumMdyn}
\end{center}
\end{figure}

We provide here a direct comparison of \Mjam\ to \Mlum, \Medyn\ and \Mhalfdyn. As defined by \citet{atlasxv} and described in section~\ref{sec:virialeqns} (equations~\ref{eqn:Mjamdefinition} and \ref{eqn:MjamMlum}) \Mjam\ is a mass parameter computed from $(M/L)_{r<\Reff}$, which provides estimates of \Mlum\ and of \Mhalfdyn. How good are these estimates? To evaluate this we can mimic the computation of \Mjam\ in our models, under the assumption that our luminous particles have a constant M/L ratio, as:
\begin{equation} 
M_{\rm JAM}  = \left( \frac{\Mdyn}{\Mlum} \right)_{r<\Reff} \, \left( \Mlum \right)_{r<r_{\rm cutoff}} .
\end{equation}
Figure~\ref{fig:MjamMlumMdyn} shows the deviations of \Mjam\ with respect to \Mlum, \Medyn\ and \Mhalfdyn, along each of the merger sequences. The figures show that \Mjam\ is an excellent measure of \Mlum\ for the initial, compact models, no matter the value of the S\'ersic index, but overestimates \Mlum\ by about 25--40 per cent once galaxies have grown at the end of the merger sequence. This is due to the increased dark matter fraction in evolved models, and in fact the deviation is about \fdm(\Reff):
\begin{equation}
\frac{\Mjam-\Mlum}{\Mlum}=\frac{\fdm(\Reff)}{1-\fdm(\Reff)} \sim \fdm(\Reff).
\end{equation}
\Mjam\ overestimates 2\,\Medyn\ by 20 per cent for $n = 4$ models, and by 40 per cent for compact $n=1$ cases. The deviation here is a function of the S\'ersic index, and is always bigger than about 20 per cent:
\begin{equation}
\frac{\Mjam - 2 \, \Medyn}{2 \, \Medyn}= \frac{\Mlum^{1/2}}{\Mlum^e} - 1 .
\end{equation}
Finally, \Mjam\ is an excellent measure of 2\,\Mhalfdyn\ in all merger stages and for all values of the initial S\'ersic index. The deviation in fact takes the form:
\[
\frac{\Mjam - 2 \, \Mhalfdyn}{2 \, \Mhalfdyn}= \frac{\fdm(\rhalflight)-\fdm(\Reff)}{1-\fdm(\Reff)} \sim 
\]
\begin{equation}
 \sim \fdm(\rhalflight)-\fdm(\Reff),
\end{equation}
which is almost always negative and smaller than 10 per cent. 

\subsection{What does $5.0\, \Reff\, \sigmae^2 / G$ measure?}
\label{sec:meaningof5sigma2Re}
From the subsections above, the expression 

\begin{equation}
\label{eqn:5sigma2Re}
M = 5.0\, \Reff\, \sigmae^2 / G
\end{equation}

\noindent may be used as an estimate of the \textit{luminous} mass \Mlum\ to an accuracy of about 30 per cent, regardless of compactness. 
That equation~ \ref{eqn:5sigma2Re} provides a good approximation to \Mlum\ was already noted by \citet{atlasxv}. 
For massive compact galaxies such as those commonly found at high redshift, equation~ \ref{eqn:5sigma2Re} needs to be corrected up by $\sim$20 per cent ($\nsers \approx 4$) to $\sim$40 per cent ($\nsers \approx 1$) in order to yield \Mlum; a much better estimate is provided by the \klum(\nsers) relation expected from equilibrium models without dark matter (Equation \ref{eq:knnodm}), with errors smaller than 10 per cent. For an extended galaxy, the dark matter fraction and the relations from Table~\ref{tab:knrel} should be considered, yielding errors of 10 -- 15 per cent from the uncertainty on the shape of the dark matter density profile. 
Since these uncertainties are significantly smaller than those of stellar masses derived from stellar population modelling techniques, equation~ \ref{eqn:5sigma2Re} provides a means of constraining the initial mass function (IMF) of galaxies from simple structural and dynamical measurements. 

How accurate is equation~ \ref{eqn:5sigma2Re} as an estimate of the dynamical mass \Mhalfdyn?  For compact galaxies, it needs to be corrected \textit{down} by $\sim$40 per cent ($\nsers \approx 4$) to $\sim$25 per cent ($\nsers \approx 1$). For extended ellipticals, equation~ \ref{eqn:5sigma2Re}  provides a good approximation to \textit{twice} \Mhalfdyn, as discussed in Sec.~\ref{sec:virialeqns}. Better estimates of \Mhalfdyn\ are provided by the \khalfdyn(\nsers) relation without dark matter (or Equation \ref{eq:knnodm} divided by 2), with errors up to 10 per cent, regardless of compactness and dark matter fraction.
Finally, equation~ \ref{eqn:5sigma2Re} provides a poor estimate of the \textit{total dynamical} mass within the luminous body of the galaxy \Mlim. Notwithstanding that our measurements of \Mlim\ are model-dependent, the predictions from equation~ \ref{eqn:5sigma2Re} need to be scaled up by factors of 2 to 3. A rough estimate of \Mlim, with accuracy of about 30 per cent, can be obtained by using $\klim = 10$ for compact galaxies and $\klim=15$ for extended ones.

\begin {table}
 \caption {\small Evolution of physical properties from the initial compact galaxy (i) to the final remnant of the dry merging process (f). } \label{tab:globalev}
\input{table6.tex}
\end{table}

\subsection{Relation between the growth in size and in velocity dispersion}
\label{sec:rhosigma}
\begin{figure}[h]
\centering
\includegraphics[scale=0.45]{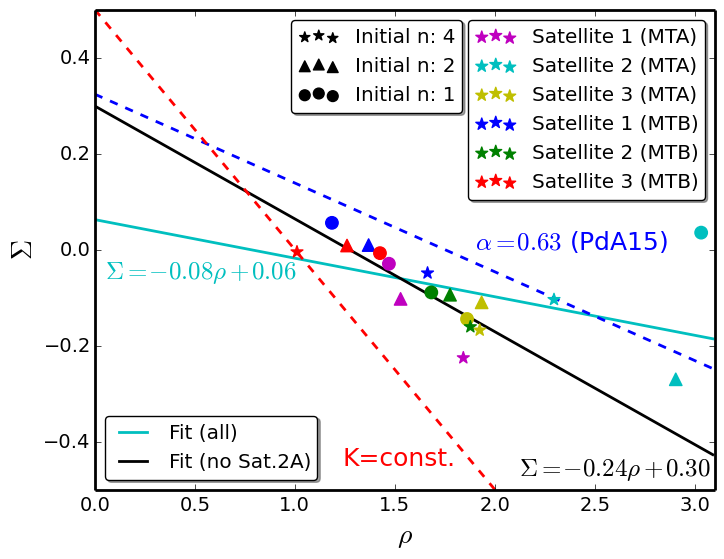}
\caption{Plot of the effective radius and velocity dispersion growth factors ($\rho$ and $\Sigma$, respectively) for every merger in our simulations.}
\label{fig:rhosigma}
\end{figure}

Several studies have shown that elliptical galaxies cannot remain homologous to themselves when undergoing mergers \citep{bertin2002, cappellari2006}. In this paper the non preservation of homology has been characterized with the S\'ersic index \nsers\ and with the virial parameters \kpar, all of which change during the mergers. Following \citet{pda2015}, another indication of non-homology can be found by comparing the increase in effective radius and effective velocity dispersion in each merger event, as a function of the corresponding increase in mass. To that end we compare the exponents $\rho$ and $\Sigma$ of the power-laws describing the evolution of \Reff\ and \sigmae\ with \Mlum, such that
\begin{equation}
\frac{\Reff^f}{\Reff^i} = \left( \frac{\Mlum^f}{\Mlum^i} \right)^\rho
\end{equation}
\begin{equation}
\frac{\sigmae^f}{\sigmae^i} = \left( \frac{\Mlum^f}{\Mlum^i} \right)^\Sigma .
\end{equation}
\noindent where superindices $i$ and $f$ denote before and after each merger, respectively. 
Assuming a constant dark matter fraction, the value of \kpar\ remains constant if $\Sigma = (1 - \rho) /2 $. However, both observations and simulations agree that galaxies evolving by mergers do not follow this relation, but rather their growth in size is not accompaigned by a corresponding decrease in velocity dispersion \citep{pda2014,pda2015}. This leads to a decrease in the value of \kpar\ during the process. In Figure \ref{fig:rhosigma} are plotted $\rho$ and $\Sigma$ for every merger presented in this paper. Our values show a trend similar to the one found in other studies, except for a slightly larger decrease in $\Sigma$. Aside from a particularly small satellite (cyan markers), the trend is fairly tight in the $\rho-\Sigma$ plane, and its best-fitting relation is:
\begin{equation}\label{eq:rhosigmatrend}
\Sigma = -0.24 \, \rho + 0.30 .
\end{equation}
The deviation from homology is thus systematic and consistent, causing a decrease in \kpar\ in every merger. In Table \ref{tab:globalev} we show the overall evolution of global parameters in each Merger Tree, together with the $\rho$ and $\Sigma$ values computed for the whole Merger Trees. Remarkably, the evolution of \klum\ seems to not depend on the initial S\'ersic index, but only on the the Merger Tree configuration; a consequence of the relation between $\rho$ and $\Sigma$ and of the weak dependence of the size growth ($\rho$) on \nsers. This consistency is then due once again to the fact that the matter from the satellites is deposited mostly in the outskirts, increasing the effective radius without touching the central regions. This process also makes the final \sigmae\ similar to what would have been the average velocity dispersion of the initial galaxy within the larger \Reff\ of the final remnant galaxy. This might explain why the relation between the evolution of \Reff\ and \sigmae\ appears so tight, and why $\Sigma$ shows a small dependence on the initial S\'ersic index. The fact that our trend shows a larger decrease in velocity dispersion than other studies might then be explained by the particularly disperse dark matter haloes; with a larger halo mass at intermediate radii the velocity dispersion wouldn't have decreased as much, and this decrease would have had an even weaker dependence on the initial S\'ersic index. In the real universe the difference in density between the progenitor and the satellites might also be smaller, allowing some satellite matter to enter the inner regions and raise the velocity dispersion. Despite all this, the trend we found can be an indication for the morphological evolution we can expect from dry mergers.

\section{Limitations}
\label{sec:shortcomings}

These results have been obtained with several assumptions which might be avoided or made better by future studies:

\begin{itemize}

\item The number of experiments is small. Only two different merger trees have been studied, hence the conclusions do not carry statistical value. As noted in the Introduction, we aim at an exploration of the trends rather than a statistical analysis to compare to observational catalogs. 

\item Initial models were spherical, non-rotating isotropic spheres. Rotation and/or dynamical anisotropy in the primary or secondary galaxies might change the structure and dynamics of the remnants. For a given galaxy mass, tangential (radial) anisotropy would cause lower (higher) velocity dispersions and thus a larger (smaller) value of \kpar. We expect these deviations to be of second order compared to the ones caused by different S\'ersic indices and dark matter fractions.

\item The size of our satellite galaxies was taken from the \citet{shen2003} relation, which made them significantly less dense than the progenitors, the size of which was picked to be very small, comparable to the observed massive compact galaxies at high redshift. Our merger experiments did not include a true major merger of equally-dense galaxies, even if we included a 1:1.4 merger in one of the merger trees. While the properties of our interacting galaxies were motivated by either observations or simulations, our setup represents only a particular case.

\item We assumed that the dark matter haloes are 10 times as massive as the luminous part of their galaxy, and we took their size from cosmological simulations, regardless of the size of the galaxy. A compact galaxy like our progenitors might however have a more dense dark matter halo, due to the deeper potential well in the centre. More precise studies on the shape of dark matter haloes are needed in order to understand this. 

\item The simulations were collisionless, while many massive compact ellipticals at high $z$ are known to contain gas \citep{papovich2006}. Gas would likely dissipate during the mergers and trigger a strong starburst in the remnant central regions, thus likely increasing the central densities \citep{barnes1996}. We expect this would contribute to `fill-in' the central density deficit seen in the surface density profiles of the $\nsers = 1$ and $\nsers = 2$ merger sequences. Gas might also have an effect at larger radii. \citet{morishita2016} suggested that gas in satellite galaxies can be triggered into star formation during otherwise `dry' mergers, contributing to the overall size increase. More in general, \citet{hopkins2009} showed that gas must be an important player in the merger-driven growth of galaxies in order to reproduce the observed morphology-mass relation, since disc galaxies would be much rarer otherwise. Nevertheless, reproducing and analysing the full variety of possible galaxy assemblies is beyond the scope of our study, which instead wanted to focus solely on the effects of dry-merger-dominated formation histories on the structure of elliptical galaxies.

\end{itemize}

\section{Conclusions}
\label{sec:conclusions}

\begin{itemize}
\item Our simulations add to the evidence that the dry merger hypothesis can explain the extreme size growth of massive elliptical galaxies between $z=2$ and $z=0$, by simulating realistic accretion histories with accurate S\'ersic-like initial models. The final remnants have an effective radius $\sim 5$ times larger than their progenitors despite having increased their mass only by a factor of $\sim 2.5$. The final effective radii, masses and dark matter fractions are compatible with galaxies in the local universe and with previous simulations on the subject. 

\item The entity of the growth induced by this process slightly depends on the S\'ersic index of the initial main galaxy: a larger $n$ corresponds to a larger overall size increase. This dependency is however small. Due to the compactness of the initial galaxy the matter from the satellites is deposited at pretty much the same radius. The dependency arises mostly because of the different amount of matter in the progenitor's outskirts, which is heated by the mergers increasing the radius of the galaxy. In either event, the dependence of the growth on the S\'ersic index of the initial main galaxy is significantly smaller than the dependence on the specific merger history of each galaxy. 

\item The growth process increases the best-fitting S\'ersic index of the main galaxy, thus breaking homology. This happens because the matter from the satellites is disproportionally deposited in the outskirts of the original galaxy. The final surface brightness profile shows deviations from the S\'ersic law; the high density contrast between primary and merging satellites contributes to the two-component nature of the surface brightness profile. The amplitude of the residuals is however not unlike that found in real ellipticals. 

\item For isolated, spherical, isotropic S\'ersic galaxies with dark matter, we provide three different parameters $\kpar \equiv G M / \Reff\, \sigmae^2$, widely used as mass estimators, and their dependency on the S\'ersic index \nsers\ and on the dark matter fraction within \Reff\ (Table~\ref{tab:knrel}; Figure~\ref{knsimul}). \klum\ and \khalfdyn\ both decrease as \nsers\ increases, whereas \klim\ increases with \nsers. Increasing the dark matter fraction, \khalfdyn\ and \klim\ increase, and \klum\ decreases. Both dependencies should be taken into account in order to have more precise mass estimates.

\item In our merger simulations \klum\  decreases, but the pattern of this decrease does not  follow the one predicted by the \kpar-\nsers\ relation of isotropic equilibrium models: \sigmae\ decreases less than what would correspond to the increase of \Reff; this shows that ellipticals grown by minor mergers are two-component, core-envelope dynamical systems not homologous to their progenitors. 
On average, the initial compact galaxies have a value of \klum\ around 7.5, the final remnants around 5. 

\item The parameter $\klim\ \equiv G \Mdyn / \Reff\, \sigmae^2$ provides the total dynamical mass within the luminous body of the galaxy from its observables \Reff\ and \sigmae. In the dry merging process its evolution is driven by the expansion of the total size of the galaxy, which encloses more and more dark matter. This dramatically raises the total dynamical mass of the galaxy within its luminous body, and thus the value of \klim.
Overall \klim\ grows from around 9 (compact galaxies at high z) to around 16 (remnants at $z=0$). 

\item The expression $M = 5\,\Reff\,\sigma^2\,/\,G$ underestimates \Mlim, the total dynamical mass within the luminous body, by a factor of $\sim$2 in $z=2$ compact ellipticals, and by a factor of $\sim$3 in $z=0$ massive ellipticals. This expression provides instead a useful approximation to the total \textit{stellar} mass of $z=0$ massive ellipticals, and thus a means of constraining the IMF of stellar populations of ellipticals from simple measurements of surface photometry and global kinematics. 

\item In each simulated merger event, the evolution \sigmae\ with \Mlum\ relates to the increase in \Reff\ with \Mlum\ in a  specific way  (Fig.~\ref{fig:rhosigma}) that is incompatible with homology: homologous evolution requires a stronger than observed decrease in \sigmae\ when \Reff\ increases. The evolution of \sigmae\ and \Reff\ is 
similar to that found in the cosmological simulations of \citet{tapia2013} and from the prepared merger simulations of \citet{hilz2012}, and is close to that inferred for real galaxies by \citet{pda2014,pda2015}. 

\end{itemize}

\section*{Acknowledgments}

\bibliographystyle{mnras}
\bibliography{mf}

\appendix

\section[]{The code}
The galaxy models used in this paper have been generated with a code written specifically for it and available to anyone interested (email M. Frigo). This code generates models of spherical galaxies with a dark matter halo, in which the surface brightness of the luminous component follows the S\'ersic law. It allows the user to choose the size and mass of both the luminous and dark components, the S\'ersic index of the luminous component, the anisotropy radii of the two components, the number of luminous/dark particles, the minimum and maximum radii of the model and the gravitational constant. The S\'ersic law tendency is achieved by adopting the Prugniel-Simien mass density profile \citep{ps1997} for the luminous component, while the dark component is generated according to the Hernquist profile \citep{hernquist1990}, which has been shown to provide a reasonably good description of dark matter haloes. The code can also generate models in which the luminous component follows the simplier Jaffe law \citep{jaffe1983}. 

\subsection{Two component Prugniel-Simien/Hernquist models}
The stability of the generated models is ensured by adopting the Eddington formula (Eq. \eqref{eq:eddi}), which given a mass density profile allows to calculate the distribution function that makes the system gravitationally stable:
\begin{equation} \label{eq:eddi}f({\cal E}) = \frac{1}{2 \sqrt{2} \, \pi^2} \left(\int_0^{\cal E} \frac{d^2\rho}{d\Psi^2} \, \frac{1}{\sqrt{\Psi-{\cal E}}} \, d\Psi + \frac{1}{\sqrt{{\cal E}}} \left(\frac{d\rho}{d\Psi} \right)_{\Psi=0} \right) \end{equation}
This equation stems from the assumption that the distribution function only depends on the energy and angular momentum of the particles, which for a stable and collisionless gravitational system is garanteed by the Jeans theorem. In a simple one-component system the potential in the second derivative is simply the one obtained by applying the Poisson equation to the mass density profile. In a system of two components however, the distribution function has to be calculated separately for each component, by using the mass density profile of the specific component and the gravitational potential of the whole system, $\Psi= \Psi_{\rm lum} + \Psi_{\rm dark}$. The code also allows to create anisotropic models, constructed according to the Osipkov-Merritt recipe (Osipkov 1977, Merritt 1985). It consists in defining a new variable 
\begin{equation} \label{eq:qpar} Q = {\cal E} + \frac{J^2}{2 r_{\rm a}}  \end{equation}
which substitutes the energy in the Eddington formula, obtaining:
\begin{equation} \label{eq:om} f(Q) = \frac{1}{2 \sqrt{2} \, \pi^2} \left(\int_0^Q \frac{d^2\rho_{\rm Q}}{d\Psi^2} \, \frac{1}{\sqrt{\Psi-Q}} \, d\Psi + \frac{1}{\sqrt{Q}} \left(\frac{d\rho_{\rm Q}}{d\Psi} \right)_{\Psi=0} \right) \end{equation}
where
\begin{equation} \label{eq:qdef} \rho_{\rm Q} (r) = \left(1+\frac{r^2}{{r_{\rm a}}^2} \right) \rho(r) .\end{equation}
The parameter $r_{\rm a}$ is known as the anisotropy radius, and the model is isotropic within it and radially anisotropic outside of it. \newline
Numerically, the code first assigns the positions of the particles in a pseudorandom way that makes them follow the given mass density profile, using the inversion sampling method. Then it assigns the velocities according to the distribution function obtained from the Eddington/Osipkov-Merritt formulas, using the acceptance-rejection method. The distribution function is calculated for each radius at the start of the routine, computing the integral numerically with the Gauss-Chebyshev quadrature.

\begin{figure}
\centering
\includegraphics[scale=0.45]{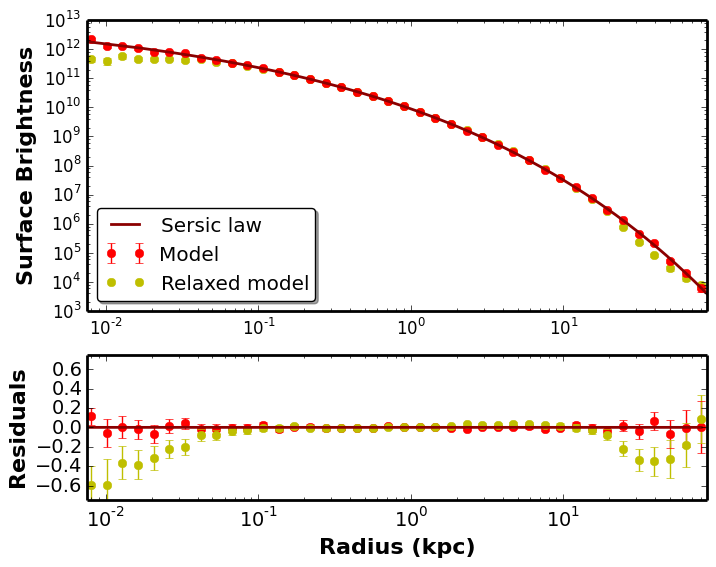}
\caption{\small Upper panel: Surface mass density profile of one of the models fitted with a S\'ersic law. The input value of \nsers\ was 4, while the best-fitting value is 4.21. Lower panel: residuals of the fit.}
\label{sersicbefore}
\end{figure}

\begin{figure}
\centering
\includegraphics[scale=0.45]{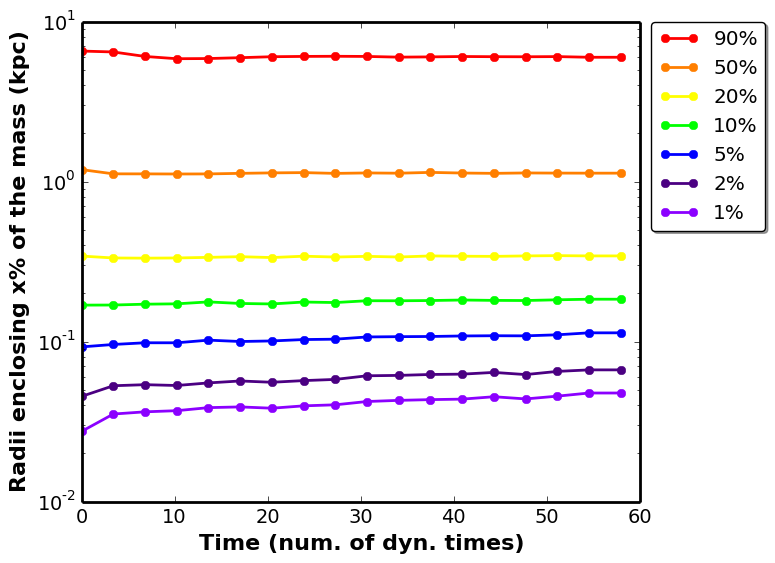}
\caption{\small Radii enclosing 1 per cent, 2 per cent,... of the luminous mass plotted over time for one of the models (progenitor galaxy of Merger Tree B).}
\label{stability}
\end{figure}

\begin{figure*}
\centering
\includegraphics[scale=0.65]{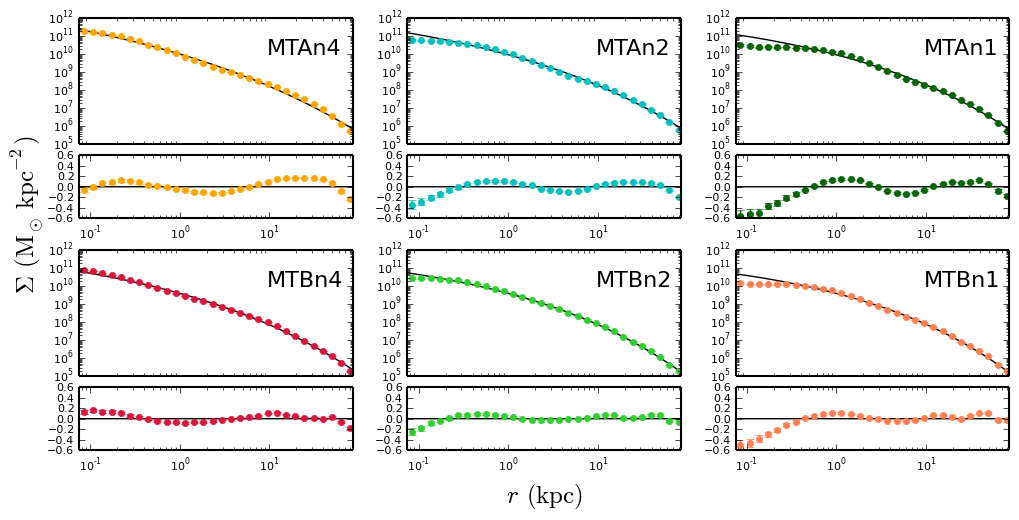}
\caption{\small S\'ersic fits of the surface mass density of all the 6 final remnants.}
\label{multisersic}
\end{figure*}

\subsection{Accuracy and stability of the generated models}
The generated models follow remarkably well the corresponding Prugniel-Simien and Hernquist profiles with the same input parameters. The surface brightness of the models thus follows a S\'ersic law, with S\'ersic index close to the one selected by the user (Fig. \ref{sersicbefore}). The stability of the models can be observed in Figure \ref{stability}, where the radii enclosing 1,2,5,... per cent of the luminous mass are plotted over time for an isolated model. In about 100 dynamical times we observe almost no variation in the half-mass radius, or even in the 10-per-cent mass radius. The 1-per-cent and 2-per-cent radii see a slight increase, due to numerical limitations (limited number of particles in the central region, softening lenght,...). The 90-per-cent radius sees a slight decrease instead, due to the fact that the model is cut after the selected maximum radius.

\bsp

\label{lastpage}

\end{document}

%% file: table3.tex
 \begin{center}
  \begin{tabular}{ | c | c | c | c | c | c | c |}
    \hline
     Merger & Stage& \Mlum & \Reff  &$\sigma_{\rm e}$ & \nsers & ${\chi_{\nu}}^2 $ \\ 
     Tree &   & $(10^{11} {\rm M}_{\odot})$ & (kpc)  & $({\rm km \, s}^{-1})$ &   &     \\ \hline \hline
    \multirow{5}{*}{$A, n=4$} & 0  & $ 1.62 \pm 0.00 $ & $ 1.06 \pm 0.00 $ & $ 328.94 \pm 0.93 $ & $4.15 \pm 0.05$ & $ 1.2 $ \\   
                              & 1  & $ 2.03 \pm 0.01 $ & $ 1.61 \pm 0.02 $ & $ 312.71 \pm 1.30 $ & $5.92 \pm 0.20$ & $ 6.6 $ \\   
                              & 2  & $ 2.17 \pm 0.01 $ & $ 1.89 \pm 0.03 $ & $ 310.33 \pm 1.02 $ & $6.81 \pm 0.23$ & $ 9.4 $ \\   
                              & 3  & $ 3.86 \pm 0.02 $ & $ 5.70 \pm 0.43 $ & $ 281.71 \pm 4.55 $ & $7.49 \pm 0.42$ & $ 103.0 $ \\   
                              & 3p & $ 1.58 \pm 0.00 $ & $ 1.11 \pm 0.01 $ & $ 315.88 \pm 1.08 $ & $4.58 \pm 0.07$ & $ 6.1 $ \\ \hline
    \multirow{5}{*}{$A, n=2$} & 0  & $ 1.66 \pm 0.00 $ & $ 1.04 \pm 0.00 $ & $ 312.73 \pm 1.10 $ & $2.06 \pm 0.02$ & $ 1.0 $ \\   
                              & 1  & $ 1.99 \pm 0.01 $ & $ 1.38 \pm 0.01 $ & $ 306.81 \pm 1.50 $ & $3.33 \pm 0.11$ & $ 31.2 $ \\   
                              & 2  & $ 2.12 \pm 0.01 $ & $ 1.64 \pm 0.01 $ & $ 301.93 \pm 0.88 $ & $3.81 \pm 0.11$ & $ 49.6 $ \\   
                              & 3  & $ 3.86 \pm 0.01 $ & $ 5.31 \pm 0.40 $ & $ 282.50 \pm 3.88 $ & $5.45 \pm 0.27$ & $ 79.3 $ \\   
                              & 3p & $ 1.64 \pm 0.00 $ & $ 1.31 \pm 0.01 $ & $ 305.25 \pm 0.68 $ & $2.10 \pm 0.03$ & $ 11.9 $ \\ \hline
    \multirow{5}{*}{$A, n=1$} & 0  & $ 1.67 \pm 0.00 $ & $ 1.03 \pm 0.00 $ & $ 309.01 \pm 0.95 $ & $1.04 \pm 0.01$ & $ 1.0 $ \\   
                              & 1  & $ 1.93 \pm 0.02 $ & $ 1.26 \pm 0.01 $ & $ 307.18 \pm 1.64 $ & $2.13 \pm 0.06$ & $ 136.5 $ \\   
                              & 2  & $ 2.08 \pm 0.01 $ & $ 1.56 \pm 0.01 $ & $ 307.40 \pm 0.68 $ & $2.68 \pm 0.07$ & $ 169.9 $ \\   
                              & 3  & $ 3.84 \pm 0.02 $ & $ 5.06 \pm 0.41 $ & $ 281.09 \pm 3.98 $ & $4.86 \pm 0.25$ & $ 160.3 $ \\   
                              & 3p & $ 1.66 \pm 0.00 $ & $ 1.50 \pm 0.01 $ & $ 305.01 \pm 1.16 $ & $1.27 \pm 0.01$ & $ 10.7 $ \\ \hline
    \multirow{5}{*}{$B, n=4$} & 0  & $ 0.50 \pm 0.00 $ & $ 0.99 \pm 0.00 $ & $ 190.32 \pm 0.70 $ & $3.89 \pm 0.05$ & $ 1.1 $ \\   
                              & 1  & $ 0.84 \pm 0.00 $ & $ 2.37 \pm 0.14 $ & $ 185.27 \pm 3.78 $ & $6.16 \pm 0.35$ & $ 6.5 $ \\   
                              & 2  & $ 0.98 \pm 0.00 $ & $ 3.20 \pm 0.15 $ & $ 180.62 \pm 2.64 $ & $6.64 \pm 0.22$ & $ 14.6 $ \\   
                              & 3  & $ 1.49 \pm 0.00 $ & $ 4.84 \pm 0.26 $ & $ 180.39 \pm 3.99 $ & $5.16 \pm 0.11$ & $ 71.0 $ \\   
                              & 3p & $ 0.47 \pm 0.00 $ & $ 1.14 \pm 0.01 $ & $ 189.19 \pm 0.83 $ & $7.04 \pm 0.21$ & $ 6.3 $ \\ \hline
    \multirow{5}{*}{$B, n=2$} & 0  & $ 0.51 \pm 0.00 $ & $ 0.97 \pm 0.00 $ & $ 181.25 \pm 0.39 $ & $1.99 \pm 0.02$ & $ 1.1 $ \\   
                              & 1  & $ 0.83 \pm 0.00 $ & $ 1.89 \pm 0.07 $ & $ 181.43 \pm 2.49 $ & $4.52 \pm 0.17$ & $ 42.1 $ \\   
                              & 2  & $ 0.97 \pm 0.00 $ & $ 2.52 \pm 0.08 $ & $ 178.81 \pm 2.69 $ & $5.04 \pm 0.15$ & $ 40.7 $ \\   
                              & 3  & $ 1.49 \pm 0.01 $ & $ 4.31 \pm 0.25 $ & $ 179.57 \pm 4.51 $ & $4.61 \pm 0.13$ & $ 29.0 $ \\   
                              & 3p & $ 0.49 \pm 0.00 $ & $ 1.10 \pm 0.01 $ & $ 179.85 \pm 0.43 $ & $2.91 \pm 0.05$ & $ 28.9 $ \\ \hline
    \multirow{5}{*}{$B, n=1$} & 0  & $ 0.52 \pm 0.00 $ & $ 0.96 \pm 0.00 $ & $ 179.74 \pm 0.73 $ & $1.03 \pm 0.01$ & $ 1.1 $ \\   
                              & 1  & $ 0.80 \pm 0.00 $ & $ 1.63 \pm 0.03 $ & $ 183.09 \pm 2.70 $ & $3.34 \pm 0.10$ & $ 154.8 $ \\   
                              & 2  & $ 0.95 \pm 0.00 $ & $ 2.22 \pm 0.05 $ & $ 180.20 \pm 2.07 $ & $4.30 \pm 0.16$ & $ 176.9 $ \\   
                              & 3  & $ 1.49 \pm 0.01 $ & $ 4.19 \pm 0.27 $ & $ 179.78 \pm 4.32 $ & $4.24 \pm 0.12$ & $ 100.8 $ \\   
                              & 3p & $ 0.50 \pm 0.00 $ & $ 1.20 \pm 0.01 $ & $ 183.01 \pm 0.54 $ & $1.49 \pm 0.02$ & $ 28.3 $ \\ \hline
    Tapia & 0   & 1.48 &  1.02 & 414.48 &  ... &  ...   \\ 
      A & 3   & 3.63 & 5.75  & 412.40 & ... & ...  \\ \hline
    Tapia & 0   & 0.45 &  0.96 & 231.35 &  ... &  ...   \\ 
      B & 3   & 1.46 & 6.74  & 275.86 & ... & ...  \\ 
    \hline
  \end{tabular}
 \end{center}

%% file: table4.tex
\begin{center}
\begin{tabular}{ | c | c | c | c |}
\hline 
$f_{\rm DM}$ & $\khalfdyn(\nsers)$  & $\klum(\nsers)$  & $\klim(\nsers)$ \\ 
\hline \hline
 0\% & $4.17 - 0.32 \, n  + 0.01 \, n^2$ & $8.34 - 0.64 \, n  + 0.01 \, n^2$ & $8.34 - 0.64 \, n  + 0.01 \, n^2$ \\ \hline
15\% & $3.56 + 0.13 \, n  - 0.05 \, n^2$ & $5.93 - 0.17 \, n  - 0.00 \, n^2$ & $6.34 + 6.15 \, n  - 0.76 \, n^2$ \\ \hline
30\% & $3.30 + 0.38 \, n  - 0.08 \, n^2$ & $4.39 + 0.06 \, n  - 0.00 \, n^2$ & $6.12 + 7.57 \, n  - 0.84 \, n^2$ \\ \hline
45\% & $3.27 + 0.46 \, n  - 0.08 \, n^2$ & $3.25 + 0.16 \, n  - 0.00 \, n^2$ & $6.36 + 7.32 \, n  - 0.72 \, n^2$ \\ \hline
75\% & $3.17 + 0.56 \, n  - 0.07 \, n^2$ & $1.76 - 0.02 \, n  + 0.02 \, n^2$ & $7.09 + 3.31 \, n  - 0.14 \, n^2$ \\ \hline
 \end{tabular}
 \end{center}

%% file: table5.tex
 \begin{center}
  \begin{tabular}{ | c | c | c | c | c | c | }
   \hline 
   ID & \khalfdyn\ & \klum\ & \klim\ & \nsers& $\fdm $ \\ 
   \hline \hline
   \multicolumn{6}{|c|}{Merger Tree A, n=4}  \\  \hline
   {\small M0} & $3.1 \pm 0.0$ & $6.1 \pm 0.0$ & $9.8 \pm 0.1$ & $4.2 \pm 0.1$ & 1.8\% \\ \hline
   {\small M1} & $2.9 \pm 0.1$ & $5.5 \pm 0.1$ & $12.4 \pm 0.2$ & $5.9 \pm 0.2$ & 3.9\% \\ \hline
   {\small M2} & $2.7 \pm 0.1$ & $5.1 \pm 0.1$ & $13.1 \pm 0.3$ & $6.8 \pm 0.3$ & 5.5\% \\ \hline
   {\small M3} & $2.5 \pm 0.2$ & $3.7 \pm 0.5$ & $12.6 \pm 1.7$ & $7.5 \pm 0.5$ & 21.3\% \\ \hline
   \multicolumn{6}{|c|}{Merger Tree A, n=2}  \\  \hline
   {\small M0} & $3.6 \pm 0.0$ & $7.0 \pm 0.1$ & $9.6 \pm 0.1$ & $2.1 \pm 0.0$ & 3.1\% \\ \hline
   {\small M1} & $3.4 \pm 0.0$ & $6.6 \pm 0.1$ & $13.1 \pm 0.2$ & $3.3 \pm 0.1$ & 5.7\% \\ \hline
   {\small M2} & $3.3 \pm 0.0$ & $6.1 \pm 0.1$ & $14.5 \pm 0.2$ & $3.8 \pm 0.1$ & 10.1\% \\ \hline
   {\small M3} & $2.8 \pm 0.3$ & $3.9 \pm 0.5$ & $14.1 \pm 2.0$ & $5.5 \pm 0.3$ & 26.5\% \\ \hline
   \multicolumn{6}{|c|}{Merger Tree A, n=1}  \\  \hline
   {\small M0} & $3.8 \pm 0.0$ & $7.3 \pm 0.1$ & $8.6 \pm 0.1$ & $1.0 \pm 0.0$ & 3.6\% \\ \hline
   {\small M1} & $3.7 \pm 0.1$ & $7.0 \pm 0.1$ & $11.8 \pm 0.2$ & $2.1 \pm 0.1$ & 8.2\% \\ \hline
   {\small M2} & $3.5 \pm 0.0$ & $6.1 \pm 0.1$ & $13.0 \pm 0.2$ & $2.7 \pm 0.1$ & 17.9\% \\ \hline
   {\small M3} & $3.0 \pm 0.3$ & $4.1 \pm 0.6$ & $14.8 \pm 2.2$ & $4.9 \pm 0.3$ & 28.1\% \\ \hline
   \multicolumn{6}{|c|}{Merger Tree B, n=4}  \\  \hline
   {\small M0} & $3.1 \pm 0.0$ & $6.0 \pm 0.1$ & $11.5 \pm 0.1$ & $3.9 \pm 0.1$ & 3.4\% \\ \hline
   {\small M1} & $2.5 \pm 0.2$ & $4.4 \pm 0.4$ & $12.0 \pm 1.2$ & $6.2 \pm 0.4$ & 11.3\% \\ \hline
   {\small M2} & $2.5 \pm 0.2$ & $4.0 \pm 0.3$ & $11.9 \pm 1.0$ & $6.6 \pm 0.3$ & 16.3\% \\ \hline
   {\small M3} & $2.7 \pm 0.2$ & $4.1 \pm 0.5$ & $12.2 \pm 1.4$ & $5.2 \pm 0.1$ & 21.2\% \\ \hline
   \multicolumn{6}{|c|}{Merger Tree B, n=2}  \\  \hline
   {\small M0} & $3.6 \pm 0.0$ & $6.9 \pm 0.1$ & $10.9 \pm 0.1$ & $2.0 \pm 0.0$ & 4.5\% \\ \hline
   {\small M1} & $3.2 \pm 0.2$ & $5.7 \pm 0.4$ & $15.3 \pm 1.0$ & $4.5 \pm 0.2$ & 9.9\% \\ \hline
   {\small M2} & $3.1 \pm 0.2$ & $5.2 \pm 0.3$ & $15.0 \pm 0.9$ & $5.0 \pm 0.2$ & 14.6\% \\ \hline
   {\small M3} & $3.0 \pm 0.3$ & $4.6 \pm 0.6$ & $14.2 \pm 1.8$ & $4.6 \pm 0.2$ & 21.4\% \\ \hline
   \multicolumn{6}{|c|}{Merger Tree B, n=1}  \\  \hline
   {\small M0} & $3.7 \pm 0.0$ & $7.1 \pm 0.1$ & $9.2 \pm 0.1$ & $1.0 \pm 0.0$ & 4.4\% \\ \hline
   {\small M1} & $3.4 \pm 0.2$ & $6.3 \pm 0.3$ & $14.9 \pm 0.8$ & $3.3 \pm 0.1$ & 11.0\% \\ \hline
   {\small M2} & $3.4 \pm 0.1$ & $5.7 \pm 0.3$ & $15.9 \pm 0.8$ & $4.3 \pm 0.2$ & 15.9\% \\ \hline
   {\small M3} & $3.2 \pm 0.3$ & $4.7 \pm 0.6$ & $14.7 \pm 2.0$ & $4.2 \pm 0.2$ & 23.5\% \\ \hline
  \end{tabular}
 \end{center}

%% file: table6.tex
\begin{center}
  \begin{tabular}{ | c | c | c | c | c | c | c | c |}
    \hline 
     M.T. & $\frac{\Mlum^f}{\Mlum^i} $ & $\frac{\Reff^f}{\Reff^i}$ & $\frac{\sigmae^f}{\sigmae^i}$ & $\frac{n_f}{n_i}$ & $\frac{K_{\star,f}}{K_{\star,i}}$ & $\rho$ & $\Sigma$  \\ 
\hline \hline
 A,n=4  & 2.39 & 5.40 & 0.86 & 1.80 & 0.60 & 1.94 & -0.18\\ \hline
 A,n=2  & 2.33 & 5.12 & 0.90 & 2.65 & 0.56 & 1.93 & -0.12\\ \hline
 A,n=1  & 2.30 & 4.93 & 0.91 & 4.68 & 0.56 & 1.91 & -0.11\\ \hline
 B,n=4  & 2.99 & 4.86 & 0.95 & 1.33 & 0.68 & 1.45 & -0.05\\ \hline
 B,n=2  & 2.90 & 4.42 & 0.99 & 2.31 & 0.67 & 1.40 & -0.01\\ \hline
 B,n=1  & 2.87 & 4.34 & 1.00 & 4.11 & 0.66 & 1.39 & 0.00\\ \hline
  \end{tabular}
\end{center}